\documentclass{aa}

\usepackage{graphicx}
\usepackage{txfonts}

\begin{document}

	\title{Herbig-Haro flows in B335
		\thanks{Based on observations made with the Nordic Optical Telescope, operated on the island of
			La Palma jointly by Denmark, Finland, Iceland, Norway, and Sweden, in the Spanish Observatorio
			del Roque de los Muchachos of the Instituto de Astrofisica de Canarias.}\fnmsep
		\thanks{This work is based in part on observations made with the Spitzer Space Telescope, which is operated
			by the Jet Propulsion Laboratory, California Institute of Technology under a contract with NASA.}\fnmsep
		\thanks{Based on observations made with the New Technology Telescope, ESO (La Silla) under programme ID 077.C-0524.}\fnmsep
		\thanks{Our NOT and NTT images are available in electronic form at the CDS via anonymous ftp to cdsarc.u-strasbg.fr (130.79.128.5)
			  or via http://cdsweb.u-strasbg.fr/cgi-bin/qcat?J/A+A/}
	}

	\author{M.~G\aa lfalk \inst{1} \and
		G.~	Olofsson \inst{1}
		}
    
	\offprints{\\M. G\aa lfalk, \email{magnusg@astro.su.se}}

	\institute{Stockholm Observatory, Sweden				
	      }
	\date{Received 01 Month 2007 / Accepted 01 Month 2007}

\abstract
{} 	
{To study the Herbig-Haro flows in the nearby dark globule B335. To find new HH objects and H$_2$ knots, make a proper motion map
of the flow activity and investigate physical properties through shock models.
\vspace{2mm}
} 	
{We have observed optical (H$\alpha$ and [SII]) and near-IR (2.12\,$\mu$m H$_2$) deep fields and taken optical spectra
using the 2.56\,m Nordic Optical Telescope, as well as a near-UV deep field ($U$ band) using the 3.58\,m NTT. In addition we present new
SPITZER\,/\,IRAC (3.5--8.0\,$\mu$m) and MIPS (24\,$\mu$m) observations. We use previous H$\alpha$ and 2.12\,$\mu$m H$_2$ observations
taken 15 and 9 years earlier to make proper motion maps. We then investigate the shock physics by matching our spectra with planar shock models.
\vspace{2mm}
}	
{We discover five new HH objects (HH\,119\,D--H) in the eastern and one (HH\,119\,I) in the western lobe of the
outflow. From proper motions we find an optically bright, roughly E--W oriented group with high space
velocities (200--280\,km\,s$^{-1}$) and a near-IR bright, slower group (15--75\,km\,s$^{-1}$) moving to
the ESE. We also find a system of at least 15 H$_2$ knots in the western lobe.
This (WNW) counterflow suggests the possibility of a binary outflow source, giving rise to two outflow axes with slightly different orientations.
We find that the E--W flow is symmetrical with evidence for two outbursts. We make the first detection of [OI]~$\lambda\lambda$~6300/63 in
HH\,119\,B and H$\beta$ in HH\,119\,A and B and find their extinctions to be A$_V$\,$\approx$\,1.4 and 4.4, respectively. HH\,119\,A is found to
expand much faster than expected from linear expansion with distance from the outflow source. Using planar shock models we find shock velocities
of $\sim$60\,km\,s$^{-1}$ (A) and $\sim$35\,km\,s$^{-1}$ (B and C). This agrees with A being of higher excitation than B and C. In our
$U$ image we detect three of the HH objects and propose that the emission arise from the [OII]\,$\lambda$3728 line and the blue continuum.
New SPITZER\,/\,IRAC and MIPS observations show most of the HH objects at 4.5\,$\mu$m and a E--W elongated hour-glass shaped structure at the
outflow source. Even at 24\,$\mu$m it is not clear whether most of the light is direct or reflected.
}
{}

\keywords{ISM: jets and outflows -- Infrared: ISM -- ISM: lines and bands -- Stars: formation -- ISM: individual objects: B335, HH119}

\maketitle

\section{Introduction}

The nearby prototypical globule B335 is located at about 250~pc, as estimated from star counts (Tomita et al. \cite{tomita}).
It is very isolated and contains a deeply embedded, low-luminosity, low-temperature outflow source (Frerking \& Langer \cite{frerking}).
The extinction towards this central outflow source is extremely large (A$_V > 320$ mag, e.g. Chandler et al. \cite{chandler}). This
explains why the central object has only been seen at far-IR (Keene et al. \cite{keene}) and longer wavelengths. It has been
observed in sub-mm (Chandler et al. \cite{chandler}) and at 3.6\,cm using the VLA with increasing
resolutions, Anglada et al. \cite{anglada} (config D, $\pm$\,9$\arcsec$), Avila et al. \cite{avila} (config C, $\pm$\,3$\arcsec$) and
Reipurth et al. \cite{reipurth02} (config A, sub-arcsec resolution). These observations have shown that the source is variable over timescales
of years (Avila et al. \cite{avila}). Reipurth et al. \cite{reipurth02} found the outflow source to be clearly elongated at 3.6\,cm in the
east-west direction, along the outflow axis, indicating a compact thermal radio jet. The combined observations makes the B335 outflow
source one of the clearest examples of a so-called Class 0 source (Andr\'{e} \cite{andre}), 
the earliest evolutionary phase of star formation, in which a deeply embedded protostar and a bipolar outflow exists.

The central $\sim$1\,$\times$\,1\,$\arcmin$ has previously been observed in deep $H$ and $K$ band images using both Keck I (Hodapp \cite{hodapp})
and the HST (Harvey et al. \cite{harvey01}). Observational evidence for current protostellar collapse also exists, based on 
millimetre wave interferometry (Harvey et al. \cite{harvey03}).

B335 is ideal for a case study of isolated star formation, as there is absolutely no source confusion caused by OB stars or outflows from other
young stars, unlike most star forming regions. Its geometry is similar to that of the well-known HH\,30 (disk and outflow), being
seen almost exactly from the side it also contains a bipolar molecular outflow accompanied by several Herbig-Haro (HH) objects lying in a line with
roughly constant declination through the deeply embedded outflow source. The axis of this HH flow coincides with that of the molecular outflow.
However, as can been seen in an optical deep field of B335 (e.g. the one in this paper) the extinction drops off rapidly about one arcminute from
the centre, making it possible to do optical spectroscopy of several of the HH objects.

The three HH objects known prior to this paper are called HH\,119\,B (discovered as a HH candidate by Vrba et al. \cite{vrba} and confirmed by
Reipurth et al. \cite{reipurth}) and HH\,119\,A and C (Reipurth et al. \cite{reipurth}). Hodapp et al. (\cite{hodapp}) discovered several
H$_2$ features (called HH\,119 IR 1--5) using the 2.12\,$\mu$m line of H$_2$. In this paper we contribute with a detailed investigation of the
B335 HH flows using mid-IR Spitzer data, deep near-UV, optical and near-IR imaging and optical spectra, revealing six new HH
objects (two with previously seen near-IR counterparts). We calculate the proper motion of HH objects
and H$_2$ features and for HH\,119\,A--C we use planar shock models to calculate shock velocities, shock compression, preshock densities and extinctions.

\section{Observations and reductions}

\subsection{Ground based}

All our ground based observations, except the near-UV imaging, were carried out at the 2.56\,m Nordic Optical Telescope (NOT) located at
2382\,m above sea level on the island of La Palma, Canary Islands. We have used the two cameras ALFOSC (Optical spectroscopy and imaging) and
NOTCam (Near-IR imaging). The near-UV deep field was obtained at the 3.58\,m New Technology Telescope (NTT) located at La Silla
Observatory, Chile. For details, see the subsections below.

\subsubsection{Optical spectra}

The ALFOSC (Andalucia Faint Object Spectrograph and Camera) long-slit spectra along the optical Herbig-Haro flow of HH\,119 in B335 were obtained
16 May 2003 using grism \#7 (3850--6850 \AA) with a slit width of 2\farcs5 (R = 260). This slit width was suitable since the centres
of HH\,119\,A--C have the same declinations within 1\farcs 3 of each other, however, HH119~A is extended but with a condensation to the south
(HH119~A$_2$; Reipurth et al. \cite{reipurth}) that was located inside the slit. With a slit length of 6\farcm5 it was possible to observe all
three objects in one slit for a total exposure time of 2400\,s. Even though the Moon was full, this did not influence the spectra noticeably since
a total lunar eclipse was in fact occurring at the time of the observations. ALFOSC has a 2048\,$\times$\,2048 array with a pixel size of 0\,\farcs188
and the seeing was about 1\farcs 0 during the observations. There were only three known HH objects (A--C) at the time of the observations, but a
fourth HH object (D), east of object C, was also within the slit due to the alignment of the slit along constant declination.

\begin{figure*}
	\centering
	\includegraphics[width=18cm]{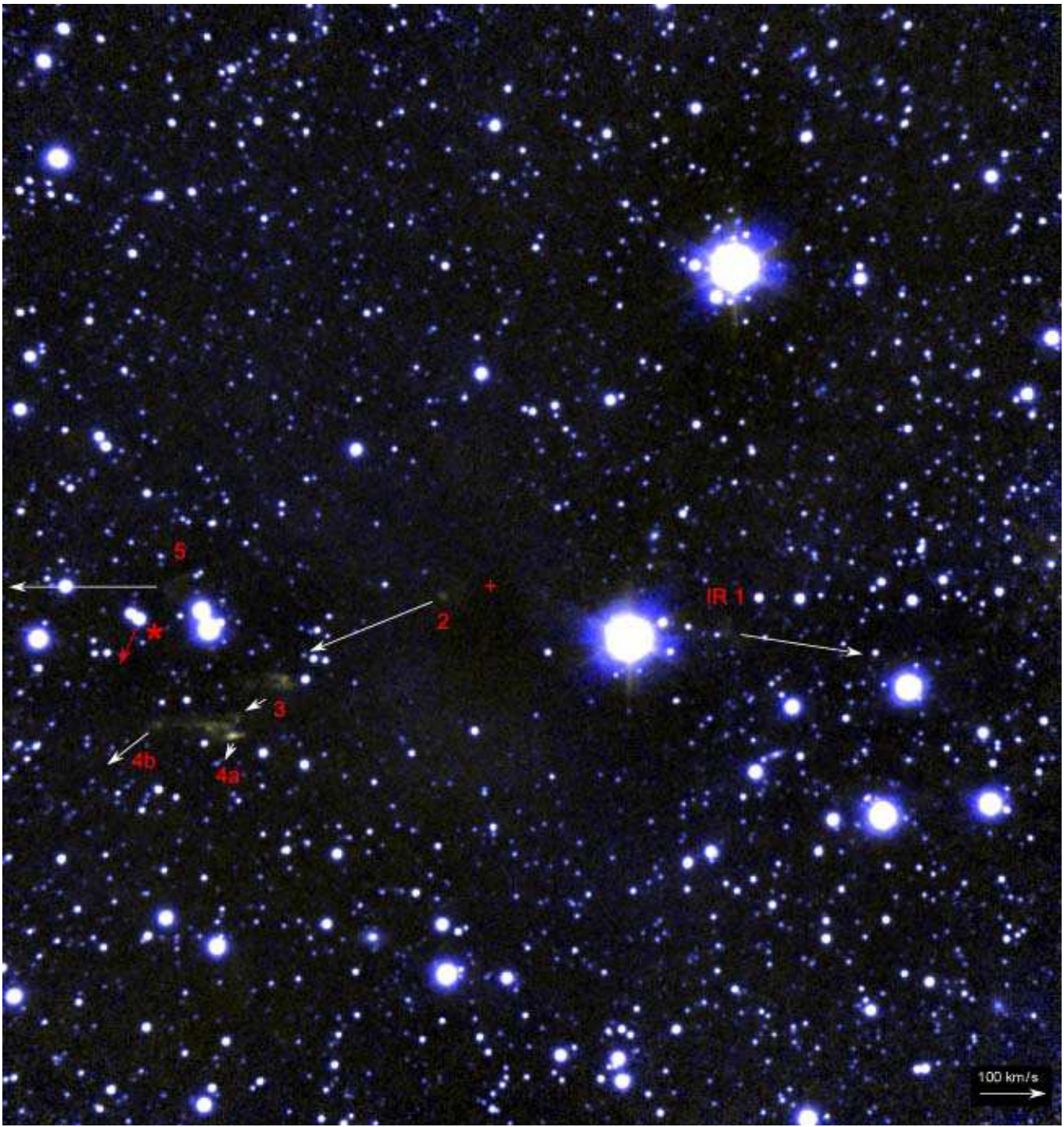}
	\caption{Near-IR deep field of B335 in the 2.12\,$\mu$m S(1) line of H$_2$ (yellow) and K$_S$ (blue) representing the continuum. All
	previously known H$_2$ features are confirmed to belong to the bipolar outflow and their proper motions are shown by the arrows.
	The red arrow shows the high proper motion of a star (asterisk).
	The position of the outflow source as seen by VLA in the 3.6\,cm continuum is marked with a plus sign. The proper motions are
	divided into two groups, HH\,119 IR\,1 and 5 moving along constant declination (similar to the optical HH flow) and
	IR\,2--4 moving more to the south east. However, the groups emanate from the same outflow source, as suggested by
	their proper motions and the cone-like feature at IR\,2, with IR\,2--4 and IR\,5 being on opposite sides of a cone shaped
	cavity created by the outflow. The field of view is 3\farcm89\,$\times$\,4\farcm13.
	}
	\label{B335_IR}
\end{figure*}

\subsubsection{Near-IR deep field (2.12\,$\mu$m H$_2$)}
\label{S1deepsec}
We have made deep near-IR observations of B335 using the 2.12\,$\mu$m $\nu=1-0$ S(1) line of H$_2$. In order to use as much telescope time as possible
for the line observations, the corresponding continuum observations were made using a $K_S$ filter (2.14\,$\mu$m central wavelength). 
The observations were made during 4 photometric nights, 4--8 July 2004, under seeing conditions that varied from poor to excellent.
We used nb exposure times of 119\,s and $K_S$ (continuum) exposures of 32\,s to avoid saturation
in most bright stars. These exposure times had been tested on a previous run and were shown to keep the near-IR camera, NOTCam, within its linear
regime throughout most of the image, without any large transient effects from bright sources that could otherwise have made a strong pattern in the
mosaic from the dithering involved.
After careful selection of all exposures, based on seeing, the final mosaic is both deep and sharp, with a mean seeing of 0\,\farcs68 and total central
mosaic exposure times of 23\,205\,s (6.45\,hours) and 2\,880\,s (48\,min) for the 2.12\,$\mu$m line and $K_S$ band respectively.

NOTCam is an HAWAII 1024\,$\times$\,1024\,$\times$18.5\,$\mu$m pixel HgCdTe array with a field-of-view of 4\farcm0\,$\times$\,4\farcm0 (pixel size of
about 0\,\farcs23). Since the observations are used for proper motion calculations we applied a distortion model for
NOTCam (G\aa lfalk \cite{notcamdist}) using only the very sharpest exposures
by mapping a relation between the on-CCD pixel positions of all known 2MASS stars in the field to their known coordinates. Using this model we have
successfully removed most of the previously seen distortion, and even at the mosaic edges the PSF of stars is circular.

A median sky is subtracted from the target (with equal exposure time) and in these observations, where we do not have any really extended objects, we use
small-step dithering between each exposure and calculate a median sky from the on-frames themselves. This effectively more than doubles the
on-target telescope time when compared to beam switching. The flatfielding is also differential (skyflats observed with some time delay and
subtracted). Besides the usual reduction steps of near-IR imaging, using in-house routines (written in IDL), we also wrote subroutines to
find and remove bad pixels, remove image distortion, sub-pixel shifting and adding overlapping images to a mosaic taking into account pixel
weighting of individual pixel exposure times. We also made a routine to remove any dark stripes that results from lowered sensitivity after a
bright source has been read out of the CCD - this could go unnoticed for normal imaging but when a deep field is made with a lot of overlaps this can
introduce a complicated pattern (especially after distortion correction) that has to be removed in order to be able to keep a high contrast throughout
the mosaic.

The two mosaics were then sub-pixel shifted in order to coincide, scaled in brightness relative to each other and a constant added in order to
emphasize H$_2$ line emission by having the same background levels in both filter mosaics and setting a suitable contrast. The final image is a
colour composite made by RGB colour coding the line image into the red and green channels and the K$_S$ (continuum) image into the blue channel,
resulting in a deep image (Fig.~\ref{B335_IR}) where H$_2$ emission features are clearly yellow while stars, galaxies and reflection nebulae are
made white/blue.


\begin{figure*}
	\centering
	\includegraphics[width=18cm]{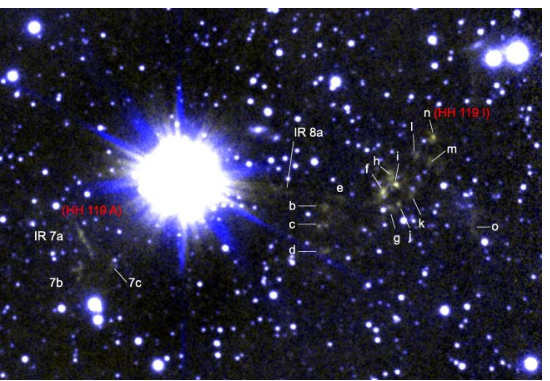}
	\caption{Near-IR image of western B335 in the 2.12\,$\mu$m S(1) line of H$_2$ (yellow) and K$_S$ (blue) representing the continuum.
	A near-IR counterpart of HH\,119\,A (which we call IR\,7) is found as well as an extended new system of S(1) features further to the west.
	The HH\,119\,A counterpart consists of three parts, two fairly bright S(1) features (IR\,7a and b) reminiscent of the optical shock but
	clearly broken up and a fainter part (7c) coinciding with a region of extended [SII] emission in the optical. We find 15 S(1) knots
	(IR\,8a--o) in the extended S(1) feature to the west, one (IR\,8n) which is also seen in H$\alpha$ and [SII] (Fig.\,\ref{New_opt_flow}) and 
	classified as an Herbig-Haro object (HH\,119\,I). The field of view is 2\farcm30\,$\times$\,1\farcm58.
	}
	\label{New_S1_flow}
\end{figure*}

\subsubsection{Western near-IR field (2.12\,$\mu$m H$_2$)}

In May 2007 we made additional S(1) observations, also using NOTCam, centred on HH\,119\,A which had not been imaged before at near-IR wavelengths.
This was also an attempt to find new H$_2$ knots in the western outflow. The observations were carried out on May 28 with total on-target exposure times
of 90 minutes in S(1) and 19.7 minutes in K$_S$. The reductions were made in a similar way as desribed in the S(1) deep field section
(cf Sect.~\ref{S1deepsec}). The seeing was excellent (0.31--0\farcs45 in K$_S$) throughout the night. Part of the resulting image can be seen in
Fig.~\ref{New_S1_flow} which shows three near-IR counterparts of HH\,119\,A and a new system of at least 15 S(1) knots to the west.


\begin{figure*}
	\centering
	\includegraphics[width=18cm]{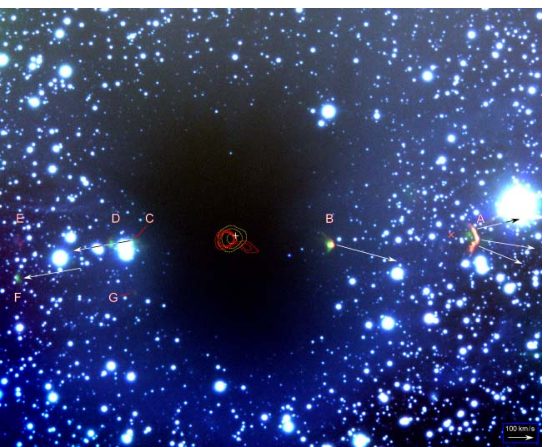}
	\caption{Optical deep field of B335 using a composite of H$\alpha$ (red), S[II] (green) and the $R$ band (blue).
	The first thing we note is that the molecular cloud surrounding the globule shines unevenly across the image, with both nebulous
	patches and hole-like features. This is probably caused by scattered starlight (in the $R$ band).
	HH\,119 A--E are located in an almost perfect line along constant declination, with proper
	motions showing them to be part of the same bipolar HH flow. Note that the proper motion arrow for object F is displaced to fit inside the
	Figure. The Spitzer contours show the outflow source region at 8.0\,$\mu$m (red, solid curves) and 24\,$\mu$m (yellow, dotted curves).
	The VLA outflow source position is marked with a
	white plus sign. At 8.0\,$\mu$m there is a clear bipolar-like feature caused by reflected light whereas the 24\,$\mu$m observations
	have a single peak with a location close to the VLA source. The red cross marks the origin HH\,119\,A {\it would have had} based on its current
	expansion alone (Fig.\ref{HH119A}), suggesting that this bow shock is currently expanding much faster than before.
	The field of view is 5\farcm18\,$\times$\,4\farcm20.
	}
	\label{B335_opt}
\end{figure*}


\begin{figure*}
	\centering
	\includegraphics[width=18cm]{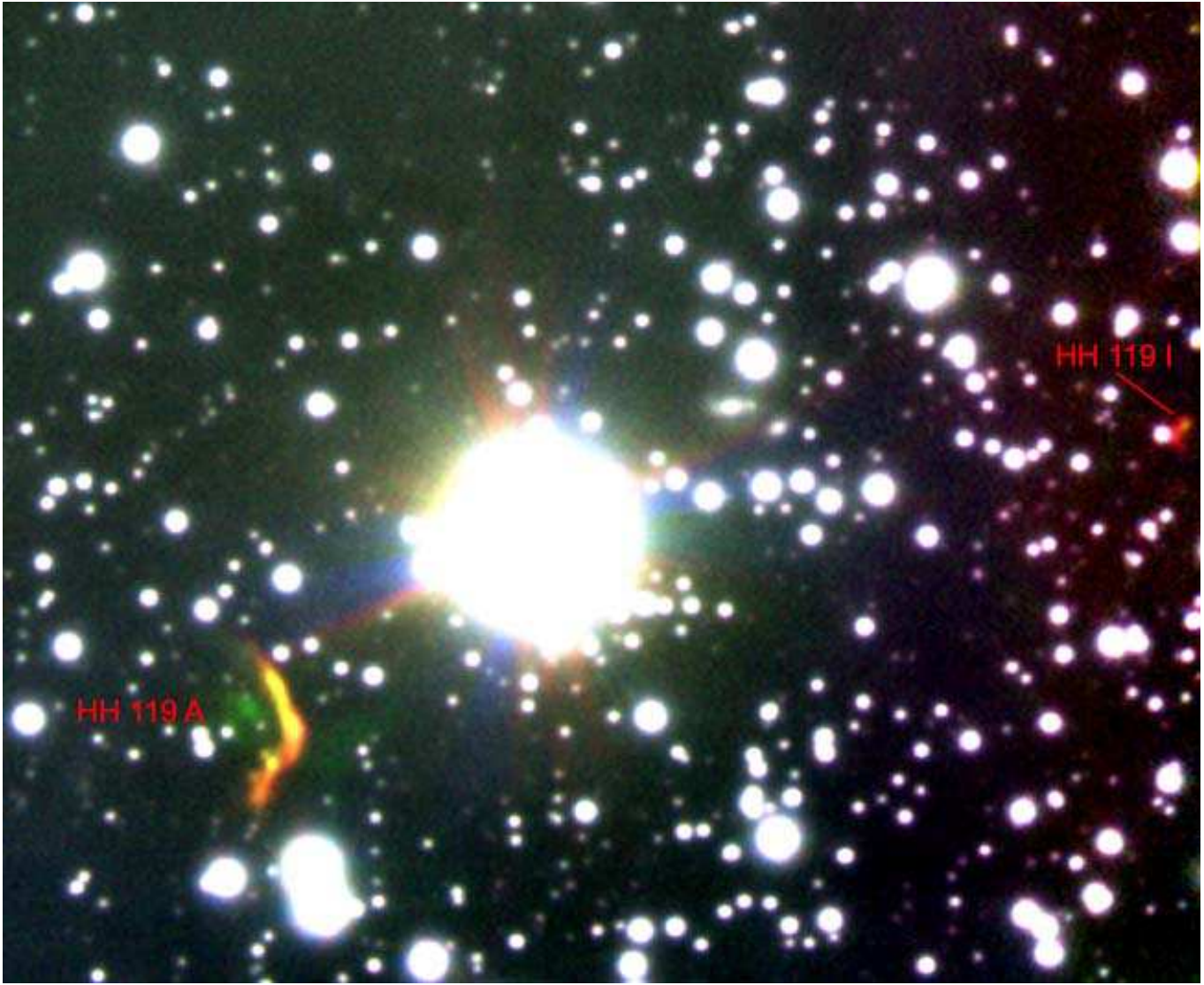}
	\caption{The westernmost part of our H$\alpha$ (red), S[II] (green) and $R$ band (blue) deep field. The number of
	exposures and therefore S/N decreases to the west because of flat field uncertainties and image dithering during the observations. After
	correcting for background variations most of the extended system of S(1) knots is covered, however, only one of these bright S(1) knots
	is detected in the optical. This object, IR\,8n, is seen in both H$\alpha$ and [SII] and confirmed as a bona-fide HH object (HH\,119\,I).
	The field of view is 1\farcm95\,$\times$\,1\farcm60.
	}
	\label{New_opt_flow}
\end{figure*}

\begin{figure*}
	\centering
	\includegraphics[width=18cm]{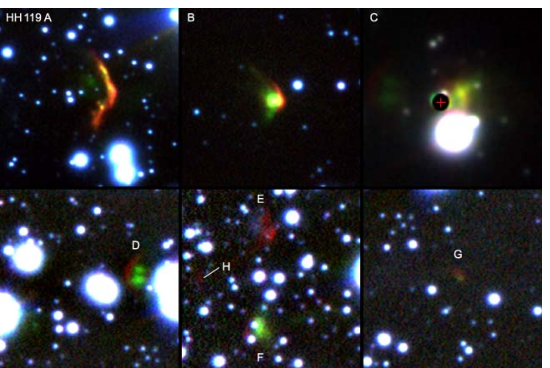}
	\caption{All Herbig-Haro objects in our optical deep field (HH119\,A--H), except object I (See Fig.\ref{New_opt_flow}).
	The colour composition is the same as in Fig.\ref{B335_opt}. HH119\,A and B belong to the west outflow lobe while C--H are part of the eastern lobe.
	In the case of object C, located very close to a bright star, this star has been removed at the position marked with a red plus sign, using
	the PSF of another bright star in the image. The central part of the removed star has been blocked out because of strong residuals
	that could not be removed. Another star can be seen very close to the north of the removed PSF and the C shock itself is fairly extended in
	both H$\alpha$ and [SII] and shows no strong condensations. Objects A, B, D and F clearly have H$\alpha$ bow-shocks leading a [SII] cooling
	zone, meaning that the flow moves into a less dense ambient medium as opposed to shock G which moves into a denser medium. Object E has
	a leading shock dominated by a line in the $R$ band that is not H$\alpha$ or [SII], probably [OI]\,$\lambda$$\lambda$\,6300/63 emission.
	The field of view in each panel is 38$\arcsec$\,$\times$\,38$\arcsec$, except for HH\,119\,C where the field size is
	24$\arcsec$\,$\times$\,24$\arcsec$.
	}
	\label{Shocks_optical}
\end{figure*}

\subsubsection{Optical deep field (H$\alpha$ and [SII])}

As a follow up to the near-IR nb-imaging we have also made deep optical nb-observations in H$\alpha$ and
[SII]\,6717/31. In addition we have made corresponding $R$ band observations to sample the continuum for both nb filters.
The observations were obtained under good seeing conditions (average seeing 0\,\farcs83) during 3 photometric nights, 3--6 Aug. 2005 with
ALFOSC (Andalucia Faint Object Spectrograph and Camera). This instrument has a $2048 \times 2048$ CCD and at a PFOV of 0\,\farcs188/pixel
it has a FOV of about 6\farcm4\,$\times$\,6\farcm4. For both nb filters we used an exposure time of 600\,s and for the $R$ band 60\,s was used
in order to avoid too much saturation of bright stars. The resulting three mosaics have total central
exposure times of 24\,000\,s (6.67\,hours) H$\alpha$, 22\,200\,s (6.17\,hours) [SII] and 2\,340\,s (39\,min) for the $R$ band (continuum).

In addition to the usual reduction steps of optical imaging (bias removal, flatfielding, cosmic ray removal) using our own set of in-house
IDL routines, we also made a camera distortion model for ALFOSC using our sharpest exposure and the positions of 163 USNO-A2 calibration
stars. After distortion removal, all exposures in each filter were sub-pixel shifted and combined, resulting in three mosaics.
The final colour composite mosaic was made with the RGB colour coding H$\alpha$ (red channel), [SII] (green channel) and $R$ band (blue channel).
Efforts were made to achieve a high contrast in the final mosaic, with the same background level in all channels outside of nebulous regions and a
colour scaling that makes most stars white, nebulae blue, H$\alpha$ and [SII] emission features clearly red and green, respectively.

The resulting colour mosiac is shown in Figures\,\ref{B335_opt} and \ref{New_opt_flow}. All the detected
Herbig-Haro objects, except HH\,119\,I (Fig\,\ref{New_opt_flow}), are shown in more detail in Fig.\,\ref{Shocks_optical}.

\begin{figure*}
	\centering
	\includegraphics[width=16cm]{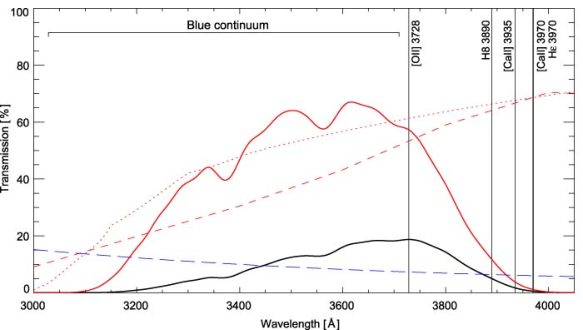}
	\caption{Transmission curves for the UV observations. The three red curves represent the three major components of the system:
	atmospheric transmission (dotted), detector QE (dashed) and the Bessel $U$ filter (solid). The solid black curve is
	the total transmission, with all three red curves included.
	Important emission lines in this wavelength range, seen in UV spectra of HH objects (Hartigan et al. \cite{hartigan99}) are marked by
	vertical lines. The horizontal line roughly marks the most contributing part of the blue continuum and the corresponding spectral energy
	distribution of the two-photon transition is shown by the blue curve (long dashed) in percentage of its peak value. 
	It is clear that only [OII]\,3728 and the blue continuum are likely contributors to the near-UV emission we see from three of the HH
	objects in B335 (A, B and E).
	}
	\label{Ufilter}
\end{figure*}

\subsubsection{Near-UV deep field}

Using EMMI on the 3.58\,m New Technology Telescope (ESO) during four nights, 28 Jun--01 Jul 2006, we have obtained 37\,$\times$\,720\,s exposures
of B335 in the $U$ band, resulting in a total exposure time of 26\,640\,s (7.4\,hours). The filter, U602, is a Bessel $U$ band filter
centred on 3542\,\AA (covering roughly 3100--4000\,\AA) and its transmission curve is shown in Fig.~\ref{Ufilter}. The usual reduction
steps of optical reductions were applied, including bias removal, flat fielding and shift-and-add to correct for tracking errors
and small-step dithering between images. The sigma-clipping technique was used to remove all cosmic rays after making the sky mean value
equal in all exposures.

The EMMI-Blue/NTT 1024\,$\times$\,1024 chip gives a pixel and image field of view
of 0\farcs362 and 6\farcm2\,$\times$\,6\farcm2, respectively. The average airmass in these observations is 1.35 and the average seeing 2\farcs03.

\begin{table*}
	\caption{Spitzer archive data used}
	\label{Spitzer}
	\begin{tabular}{clcccc}
	\hline
	\noalign{\vspace{0.5mm}}
	Prog.ID & P.I. & Key & Type & Released & Scheduled \\
	\noalign{\vspace{0.5mm}}
	\hline
	94 & Lawrence C.& 4926464 & IRAC Mapping & 2005-06-24 & 2004-04-21 \\
	53 & Rieke G. & 12022016 & MIPS Scan Map & 2005-11-09 & 2004-10-15 \\
	\hline
	\end{tabular}
\end{table*}

\subsection{Spitzer Space Telescope}

Spitzer is an IR space telescope that carries a 85\,centimeter cryogenic telescope and three cryogenically cooled science instruments, one
of these is the Infrared Array Camera (IRAC) that provides simultaneous 5\farcm2\,$\times$5\farcm2 images in four
channels, centred at 3.6, 4.5, 5.8 and 8.0 $\mu$m. Each channel is equipped with a 256\,$\times$\,256 pixel detector array with a pixel size
of about 1\farcs2\,$\times$\,1\farcs2. Another instrument is the Multiband Imaging Photometer for Spitzer (MIPS) that contains
three separate detector arrays, making simultaneous observations possible at 24, 70, and 160 $\mu$m.

The Spitzer data used in this paper (see Table~\ref{Spitzer}) were obtained from the Spitzer Science Archive
using the Leopard software. All data had been reduced to the Post-Basic Calibrated Data (pbcd) level. We use all channels of the IRAC
observations, and some in-house routines to reduce them further and especially to make a colour composite image. From the MIPS data, we
only use the 24\,$\mu$m observations because of its higher resolution and because the other two channels are full of artefacts (probably
due to saturation). The 24\,$\mu$m camera has a format of 128$\times$128 pixels and a pixel size of 2\farcs55. In order to plot Spitzer
contours on our ground-based observations, although very different resolutions, we matched all stars seen in the Spitzer 24\,$\mu$m mosaic with
both our $K_S$ and $R$ band observations and corrected for differences in FOV, distortion and image rotation.

\subsection{Proper motions}
\label{firstepoch}

\subsubsection{Near-IR first epoch image}
In order for us to map the IR proper motions of as many H$_2$ features as possible we needed a previous epoch of 2.12\,$\mu$m observations, having
a long enough time span to show the proper motions with reasonable accuracy, but at the same time with high enough quality to show as many
of our recently observed H$_2$ objects as possible given that the new epoch is a deep field. We would like to thank Klaus-Werner Hodapp for
kindly providing us with the first epoch observations, in the form of a deep (8200\,s or about 2.3 hours) mosaic taken with the
2.2\,m University of Hawaii telescope. These observations were made using the Quick Infrared Camera (QUIRC) mounted at the f/10 focus, yielding
a pixel size of 0\,\farcs185 (not that far from our second epoch images, 0\,\farcs23). Since these observations were obtained
between 8-10 July 1995 (and our second epoch observations 4-8 July 2004) we have a time span of about 9 years between the epochs. For more information
about the first epoch observations see Hodapp et al. \cite{hodapp}.

\subsubsection{Optical first epoch image}
For our optical proper motions, we used an even larger time span (roughly 15\,years) than for the IR case. We would like to thank Bo Reipurth for
kindly providing us with the original H$\alpha$ images from Reipurth et al. \cite{reipurth}, taken in August 1990 with the 3.58\,m New Technology Telescope (NTT). These observations are deep enough and have a long enough time span when combined with our images to yield accurate proper motions. The instrument
used in these observations, EMMI, provided a field size of 7\farcm5\,$\times$\,7\farcm5 when used with a Thompson 1024\,$\times$\,1024 pixel chip, yielding
a pixel size of 0\farcs44 (rougly twice the pixel size of our Aug 2005 observations).

\subsubsection{Calculations and accuracy}
We refer to Section 3 in G\aa lfalk \& Olofsson \cite{galfalk06} for details on our method for calculating proper motions. In short, we fit all stars seen in both epoch images using a two-dimensional elliptical Gaussian. Stars with measurable proper motions are eliminated in the solution.
After warping the lower resolution image to fit the one with higher resolution, the
images are shifted for each object in sub-pixel (0.1 pixel) steps until the object overlaps itself in both epochs, and a proper motion is found. For each step the matching is judged by blinking the two images at different rates and by making a colour image using two colour channels (red and blue).
The accuracy is similar for our optical and near-IR proper motions (about 12 km/s), except for HH\,119\,C (close to a bright star) and
HH\,119\,F (faint in the first epoch image).


\begin{table*}
	\caption{{\bf Geometry of all Herbig-Haro objects and H$_2$ knots.}
	}
	\label{shocksum}
	\begin{tabular}{lcccccll}
	  \hline
        \noalign{\vspace{0.5mm}}
        HH\,119\,$^{\mathrm a}$ & RA\,$^{\mathrm b}$ & Dec\,$^{\mathrm b}$ & Size\,$^{\mathrm c}$ & Regime & Structure\,$^{\mathrm d}$ & Reference & Comment \\
             & (2000) & (2000) & ($\arcsec$) & & & (first) & \\
 	\noalign{\vspace{0.5mm}}
        \hline

	\noalign{\vspace{1.0mm}}

	A (IR\,7) & 19:36:51.32 & 07:34:10.5 & 6.1 $\times$ 17.2 & Optical & H$\alpha$ / [SII] & Reipurth 1992 & Bow shock \\
	(A$_1$) & 19:36:51.32 & 07:34:10.5 & .... & " & .... & ~~~~~~~~" & H$\alpha$ peak and background star\\
	(A$_2$) & 19:36:51.47 & 07:34:06.8 & .... & " & .... & ~~~~~~~~" & H$\alpha$ peak (South) \\
	(A$_3$) & 19:36:51.46 & 07:34:15.2 & .... & " & .... & ~~~~~~~~" & H$\alpha$ peak (North) \\
	B (IR\,1) & 19:36:56.98 & 07:34:06.0 & 8.0 $\times$ 13.2 & Optical & H$\alpha$ / [SII] & Vrba 1986 & Bow shock \\
	C & 19:37:05.00 & 07:34:06.0 & 1.7 $\times$ 4.4  & Optical & H$\alpha$ + [SII] & Reipurth 1992 & No separation between [SII] and H$\alpha$ \\
	D (IR\,5) & 19:37:05.82 & 07:34:07.2 & 3.7 $\times$ 7.2  & Optical & H$\alpha$ / [SII] & This paper & H$\alpha$ bow shock, two [SII] bullets \\
	E & 19:37:09.19 & 07:34:06.5 & 5.6 $\times$ 8.6  & Optical & [SII]\,$^{\mathrm e}$ / H$\alpha$ & This paper & Leading feature bright in $R$ band$^{\mathrm e}$ \\
	F & 19:37:09.39 & 07:33:46.0 & 4.6 $\times$ 5.1  & Optical & H$\alpha$ / [SII] & This paper & Bow shock \\
	G (IR\,4a) & 19:37:04.75 & 07:33:38.3 & 2.7 $\times$ 2.9  & Optical & [SII] / H$\alpha$ & This paper & \\
	H & 19:37:10.22 & 07:33:56.8 & 5.4 $\times$ 4.4  & Optical & H$\alpha$ & This paper & More extended at 4.5\,$\mu$m \\
	I (IR\,8n) & 19:36:45.41 & 07:34:40.2 & 2.3 $\times$ 3.4 & Optical & H$\alpha$ + [SII] & This paper & Part of a WNW flow \\
	IR\,1 & 19:36:57.22 & 07:34:01.7 & 3.0 $\times$ 5.4  & Near-IR & & Hodapp 1998 & IR counterpart of HH\,119\,B \\
	IR\,2 & 19:37:01.50 & 07:34:08.5 & 2.3 $\times$ 2.4  & Near-IR & & Hodapp 1998 & \\
	IR\,3 & 19:37:03.89 & 07:33:49.6 & 11.7 $\times$ 3.5  & Near-IR & & Hodapp 1998 & \\
	IR\,4 & 19:37:05.22 & 07:33:39.3 & 21.6 $\times$ 6.3  & Near-IR & & Hodapp 1998 & Large and irregular\\
	IR\,4a & 19:37:04.66 & 07:33:36.6 & 4.5 $\times$ 1.2  & Near-IR & & This paper & IR counterpart of HH\,119\,G \\
	IR\,4b & 19:37:05.77 & 07:33:39.3 & 3.7 $\times$ 2.1  & Near-IR & & This paper & Condensation within IR\,4 \\
	IR\,5 & 19:37:05.70 & 07:34:09.5 & 13.8 $\times$ 11.5  & Near-IR & & Hodapp 1998 & IR counterpart of HH\,119\,D \\
	IR\,7a & 19:36:51.39 & 07:34:12.7 & 3.7 $\times$ 8.4 & Near-IR & & This paper & North IR segment of A \\
	IR\,7b & 19:36:51.53 & 07:34:05.4 & 2.8 $\times$ 4.7 & Near-IR & & This paper & South IR segment of A \\
	IR\,7c & 19:36:50.93 & 07:34:06.5 & 2.6 $\times$ 3.8 & Near-IR & & This paper & Segment in front of bow A \\
	IR\,8a--o & 19:36:46 & 07:34:26 & 58 $\times$ 33  & Near-IR & & This paper & Extended system of 15 H$_2$ knots \\
	IR\,8f & 19:36:46.23 & 07:34:27.2 & 0.9 $\times$ 1.6 & Near-IR & & This paper & One of the brightest knots in IR\,8 \\
	IR\,8i & 19:36:46.02 & 07:34:29.1 & 0.9 $\times$ 0.7 & Near-IR & & This paper & One of the brightest knots in IR\,8 \\
	IR\,8n & 19:36:45.39 & 07:34:41.4 & 2.3 $\times$ 2.6 & Near-IR & & This paper & IR counterpart of HH\,119\,I \\

	\noalign{\vspace{0.5mm}}
	\hline

	\end{tabular}

	\begin{list}{}{}
		\item[$^{\mathrm{a}}$] Letters designate Herbig-Haro objects, while IR numbers designate H$_2$ knots.
		\item[$^{\mathrm{b}}$] Optical positions in H$\alpha$ (04 Aug 2005) and near-IR positions in 2.12\,$\mu$m H$_2$ (06 Jul 2004 ; 29 May 2007 for IR\,7 and 8).
		\item[$^{\mathrm{c}}$] Approximate bounding box sizes, determined visually.
		\item[$^{\mathrm{d}}$] Leading / trailing flux peak.
		\item[$^{\mathrm{e}}$] The leading feature is easier seen in $R$ than [SII], most likely due to a strong line other than H$\alpha$ and [SII].
	\end{list}

\end{table*}

\begin{table*}
	\caption{{\bf Flux summary.}
	}
	\begin{tabular}{lcccccc}
	\hline
        \noalign{\vspace{0.5mm}}
        HH\,119 & F$_U$$^{\mathrm a}$ & F [H$\alpha$]$^{\mathrm b}$ & F [SII]$^{\mathrm c}$ & F [S(1)]$^{\mathrm d}$ & F$_{4.5}$$^{\mathrm e}$ & F [SII] / F [H$\alpha$]$^{\mathrm f}$ \\
	\noalign{\vspace{0.5mm}}
 	(1) & (2) & (3) & (4) & (5) & (6) & (7) \\
 	\noalign{\vspace{0.5mm}}
        \hline

	\noalign{\vspace{1.0mm}}

	A (IR\,7) & 8.50 $\pm$ 0.34 & 	17.34 $\pm$ 0.91 & 9.94 $\pm$ 0.47 & 17.12 $\pm$ 0.38 & ~~93 $\pm$ 19 & 0.573 $\pm$ 0.018 \\
	B (IR\,1) & 1.45 $\pm$ 0.07 &	05.92 $\pm$ 0.34 & 7.87 $\pm$ 0.44 & 2.73 $\pm$ 0.11 & 103 $\pm$ 13 & 1.330 $\pm$ 0.030 \\
	C & ... & 03.54 $\pm$ 0.18 & 7.36 $\pm$ 0.41 & ... & & 2.079 $\pm$ 0.018 \\
	D (IR\,5) & ... & 01.40 $\pm$ 0.08 & 2.45 $\pm$ 0.10 & 10.95 $\pm$ 0.17 & 107 $\pm$ 19 & 1.749 $\pm$ 0.077 \\
	E & 1.23 $\pm$ 0.06 &	00.79 $\pm$ 0.06 & 0.51 $\pm$ 0.03 & out & 40 $\pm$ 7 & 0.65 $\pm$ 0.11 \\
	F & ... & 00.78 $\pm$ 0.05 & 2.16 $\pm$ 0.12 & out & 67 $\pm$ 8 & 2.78 $\pm$ 0.10 \\
	G (IR\,4a) & ... & 0.144 $\pm$ 0.007 & 0.065 $\pm$ 0.004 & 6.91 $\pm$ 0.14 & ~~45 $\pm$ 11 & 0.45 $\pm$ 0.15 \\
	H (IR\,6) & ... & 00.07 $\pm$ 0.01 & ... & out & 147 $\pm$ 17 & ... \\
	I (IR\,8n) & out & 01.64 $\pm$ 0.06 & 0.24 $\pm$ 0.03 & 5.58 $\pm$ 0.18 & out & 0.148 $\pm$ 0.021 \\
	IR\,2 & ... & ... & ... & 1.36 $\pm$ 0.03 & & ... \\
	IR\,3 & ... & ... & ... & 7.71 $\pm$ 0.15 & 41 $\pm$ 6 & ... \\
	IR\,4 & ... & ... & ... & 21.59 $\pm$ 0.30 & 134 $\pm$ 13 & ... \\
	IR\,4b & ... & ... & ... & 2.40 $\pm$ 0.05 & 14 $\pm$ 3 & ... \\
	IR\,7a & ... & ... & ... & 11.04 $\pm$ 0.34 & ... & ... \\
	IR\,7b & ... & ... & ... & 5.43 $\pm$ 0.17 & ... & ... \\
	IR\,7c & ... & ... & ... & 0.66 $\pm$ 0.03 & ... & ... \\
	IR\,8f & out & ... & ... & 5.29 $\pm$ 0.17 & out & ... \\
	IR\,8i & out & ... & ... & 4.34 $\pm$ 0.14 & out & ... \\

	\noalign{\vspace{0.5mm}}
        \hline

	\end{tabular}

	\begin{list}{}{}
		\item[$^{\mathrm{a}}$] Integrated U band flux in units of 10$^{-15}$\,erg\,s$^{-1}$\,cm$^{-2}$, effective filter width is 408.8\,\AA (from total transmission curve).
		\item[$^{\mathrm{b}}$] Emergent flux in H$\alpha$, in units of 10$^{-15}$\,erg\,s$^{-1}$\,cm$^{-2}$.
		\item[$^{\mathrm{c}}$] Emergent flux in the 6717+6731 [SII] lines, in units of 10$^{-15}$\,erg\,s$^{-1}$\,cm$^{-2}$.
		\item[$^{\mathrm{d}}$] Emergent flux in the H$_2$\,$\mu$ = 1--0 S(1) 2.12\,$\mu$m line, in units of 10$^{-15}$\,erg\,s$^{-1}$\,cm$^{-2}$.
		\item[$^{\mathrm{e}}$] Integrated flux of Spitzer IRAC 2 (4.5\,$\mu$m) in units of 10$^{-15}$\,erg\,s$^{-1}$\,cm$^{-2}$. HH\,119\,C and IR2 close to bright sources.
		\item[$^{\mathrm{f}}$] Mean flux ratio including full object extension.
	\end{list}
	\label{HH_fluxtable}
\end{table*}

\section{Results and discussion}

\subsection{New HH objects and H$_2$ knots}

Using our deep optical and near-IR mosaics, in combination with the proper motions obtained from these and the additional information from optical
spectra, we confirm that two of the previously known 2.12\,$\mu$m H$_2$ features (IR\,4 and 5 in Hodapp et al. \cite{hodapp}) have optical
counterparts that are bona-fide HH objects (named HH\,119\,G and D). Four more HH objects are also discovered in our optical deep
field (we call these HH\,119\,E, F, H and I). These six new HH objects can be seen in Figures\,\ref{B335_opt}, \ref{New_opt_flow}
and \ref{Shocks_optical}. HH\,119\,H was first found as an IR H$_2$ feature in the Spitzer/IRAC mosaic (Fig.\ref{B335_IRAC}). However, it is also
faintly detected in H$\alpha$ (Fig.\ref{Shocks_optical}) and given its position in the E--W HH flow it is classified as an HH object.

Many of the HH objects have the appearance of bow shocks with clearly separated H$\alpha$ and [SII] emission. A summary of all shocks is given
in Table\,\ref{shocksum} and the photometry is presented in Table~\ref{HH_fluxtable}.
Only three of these objects, HH\,119\,A--C, were known HH objects prior to this survey. We have followed the same naming
scheme, and named Herbig-Haro objects using letters from D to I.

There is a rough symmetry in HH\,119\,A--F and H, lying approximately in a line through the outflow source, all at two roughly symmetric projected
distances, possibly corresponding to two outburst episodes in the central source.
The innermost HH object in the western lobe, HH\,119\,B (58\farcs2 from the outflow source), has two counterparts in the eastern
lobe, HH\,119\,C (61\farcs2) and D (73\farcs3). The outermost western object, HH\,119\,A (142\farcs3), has three counterparts
in the eastern lobe, HH\,119\,E (123\farcs4), F (126\farcs4) and H (139\farcs4).
HH\,119\,G is not located on this line roughly along constant declination, instead this object is moving more to the south east, in the
same group as IR knots 2, 3, 4a and 4b (Fig.~\ref{B335_IR}). All these IR knots are likely to have optical counterparts, however, due to the
high extinction in this part of the flow only knot 4a is detected in the optical and confirmed as an HH object.

In the western lobe we discover a system of 15 new H$_2$ knots and three counterparts to HH\,119\,A (Fig.~\ref{New_S1_flow}).


\begin{table*}
	\caption{{\bf IR proper motion results.} Column 1 gives the H$_2$ feature designation, as given in
	Figure\,\ref{B335_IR}. Columns 2--4 give the angular distance
	each shock has moved between the two epochs (9 years apart). Columns 5-7 are the proper
	motions and finally Columns 8--9 the (tangential) velocities and positional
	angles. Uncertainties are given at the bottom of the Table (except for the PA
	which is given for each row).
	}
	\label{propermot_IR}
	\begin{tabular}{lrrrrrrrll}
	  \hline
        \noalign{\vspace{0.5mm}}
        HH\,119 & $\Delta_{\alpha}$ & $\Delta_{\delta}$ & $\Delta_{tot}$ & $\mu_{\alpha}$ & $\mu_{\delta}$ & $\mu_{tot}$ & vel & PA\,$^{\mathrm a}$ & Comment \\
             & (\arcsec) & (\arcsec) & (\arcsec) & (mas/yr) & (mas/yr) & (mas/yr) & (km/s) & (deg) & \\
	\noalign{\vspace{0.5mm}}
 	(1)  & (2)	 & (3)	     & (4)	 & (5)	    & (6)      & (7)	  & (8)	   & (9)   & (10) \\
 	\noalign{\vspace{0.5mm}}
        \hline

	\noalign{\vspace{1.0mm}}

	IR\,1 (B) &	-1.51 &	-0.24 &	1.53 & -168 & -27 & 170 & 201 & 261 $\pm$ 3 & \\
	IR\,2 &	+1.53 &	-0.61 &	1.65 & +170 & -68 & 183 & 217 & 112 $\pm$ 3 & \\
	IR\,3 &	+0.17 &	-0.09 &	0.19 & +19 & -10 & 21 &	25  & 119 $\pm$ 24 & \\
	IR\,4a (G) &	+0.06 &	-0.11 &	0.12 & +6 & -12 & 14 &	16 & 153 $\pm$  37 & \\
	IR\,4b &	+0.45 &	-0.35 &	0.57 & +50 & -39 & 63 &	75 & 128 $\pm$  9 & \\
	IR\,5 (D) &	+1.86 &	0 &	1.86 & +207 & 0 & 207 &	245 & 90 $\pm$  2 & \\

	\noalign{\vspace{1.0mm}}

	Star &	+0.17 &	-0.35 &	0.39 & +19 & -39 & 44 &	... & 155 $\pm$ 8 & Marked in Fig.\,\ref{B335_IR} \\

	\noalign{\vspace{1.0mm}}
        \hline
        \noalign{\vspace{0.5mm}}
	   &  $\pm$ 0.06 & $\pm$ 0.06 & $\pm$ 0.09 & $\pm$ 7 & $\pm$ 7 & $\pm$ 10 & $\pm$ 12 & & \\
	\noalign{\vspace{0.5mm}}
	\hline

	\end{tabular}

	\begin{list}{}{}
		\item[$^{\mathrm{a}}$] Position Angle: North 0\,$\degr$, East 90\,$\degr$, South =180\,$\degr$, West 270\,$\degr$.
	\end{list}

\end{table*}

\subsection{Near-IR shocks}

Our 2.12\,$\mu$m H$_2$ deep field is presented in Fig.\ref{B335_IR} along with shock names and arrows illustrating the calculated
proper motions. The near-IR proper motions and projected shock velocities are also given in Table\,\ref{propermot_IR}. There is an apparent
division of the shocks into two groups, one (IR\,1 and 5) that coincides with the optical HH flow along almost constant declination, and another
group (IR\,2--4) with a position angle of about 115$\degr$ (moving to the ESE). The projected velocity
vectors suggest that both groups emanate from the outflow source in the centre of B335 (marked by a plus sign in the deep field image).
Looking at the east (blue-shifted) lobe, the faint cone-like feature close to the outflow source in the near-IR (Fig.\ref{B335_IR}) and
the mid-IR Spitzer data (Fig.\ref{B335_IRAC}), combined with the proper motions suggest that the two groups in the eastern lobe are on the opposite
sides of a cone-shaped cavity, cleared out by the molecular outflow.

While the first group (IR\,1 and 5) is bright in the optical lines, only the brightest peak in the second group, IR\,4a, is seen in our optical
H$\alpha$ and [SII] deep fields. The reason for this could be higher extinction but another reason is hinted by the slower projected
velocities, slow moving (C-type) shocks are non-dissociative, instead molecules like H$_2$ are kept intact and is a good tracer of such shock
environments. Fast moving (J-type) shocks instead break up molecules, making such shocks bright in atomic lines such
as H$\alpha$ and $\lambda$$\lambda$\,6717,\,6731~[SII].

West of the outflow source (in the red-shifted lobe) we find at least 16 H$_2$ knots, one in the deep field (IR\,1) and an extended system of
knots in the western field (Fig\,\ref{New_S1_flow}). Although we lack proper motions for this system, they are most likely part of a counterflow
to the ESE flow. This suggests the possibility of a binary outflow source, giving rise to both an E--W and an ESE--WNW outflow axis.

The only western knot for which we have previous observations, IR\,1, shows a proper motion that points almost exactly away from
the VLA outflow source. Although very different in appearance, this shock coincides very well with the southern part of the optically (H$\alpha$ and
[SII]) strong, nicely defined, bow shock HH\,119\,B. With both the near-IR and optical shocks having a proper motion position angle of
roughly 260$\degr$ and partly overlapping each other, it is clear that they are both part of the same HH flow and originate from the central outflow
source. However, how they are physically related is not equally clear from positions and proper motions alone.
A similar separation applies for the optical and near-IR images of HH\,119\,D (IR\,5). A more detailed discussion in given in Section \ref{HH119shocks}
where it is shown that both IR\,1 and IR\,5 likely trace parts of bow shocks HH\,119\,B and D, respectively, that have lower shock velocities (i.e.
the outer wings of these bow shocks).


\begin{table*}
	\caption{{\bf Optical proper motion results.} Column 1 are the HH designations, as given in
	Figure\,\ref{B335_opt}. Columns 2--4 give the angular distance
	each shock has moved between the two epochs (15 years apart). Columns 5-7 are the proper
	motions and finally Columns 8--9 the (tangential) velocities and positional
	angles. Uncertainties are given at the bottom of the table (except for the PA
	where this is given in each row).
	}
	\label{propermot_opt}
	\begin{tabular}{ccccccccc}
	  \hline
        \noalign{\vspace{0.5mm}}
        HH\,119 & $\Delta_{\alpha}$ & $\Delta_{\delta}$ & $\Delta_{tot}$ & $\mu_{\alpha}$ & $\mu_{\delta}$ & $\mu_{tot}$ & vel & PA\,$^{\mathrm a}$ \\
              & (\arcsec) & (\arcsec) & (\arcsec) & (mas/yr) & (mas/yr) & (mas/yr) & (km/s) & (deg) \\
	\noalign{\vspace{0.5mm}}
 	(1)  & (2)	 & (3)	     & (4)	 & (5)	    & (6)      & (7)	  & (8)	   & (9)   \\
 	\noalign{\vspace{0.5mm}}
        \hline

	\noalign{\vspace{1.0mm}}

	A$_1$ &	--3.00 $\pm$ 0.10 & --0.49 $\pm$ 0.10 & 3.04 $\pm$ 0.12 & --201 $\pm$ 7 & --33 $\pm$ 7 & 203 $\pm$ 9 & 241 $\pm$ 10 & 261 $\pm$ 3 \\
	A$_2$ &	--2.43 $\pm$ 0.10 & --0.97 $\pm$ 0.10 & 2.62 $\pm$ 0.13 & --162 $\pm$ 7 & --65 $\pm$ 7 & 175 $\pm$ 10 & 207 $\pm$ 11 & 248 $\pm$ 3 \\
	A$_3$ &	--2.34 $\pm$ 0.10 & +0.53 $\pm$ 0.10 & 2.40 $\pm$ 0.12 & --156 $\pm$ 7 & +36 $\pm$ 7 & 160 $\pm$ 9  & 190 $\pm$ 10 & 283 $\pm$ 3 \\
	B &	--3.40 $\pm$ 0.10 & --0.91 $\pm$ 0.10 & 3.52 $\pm$ 0.13 & --227 $\pm$ 7 & --61 $\pm$ 7 & 235 $\pm$ 9 & 278 $\pm$ 11 & 255 $\pm$ 3 \\
	C &	+3.42 $\pm$ 0.19 & --0.57 $\pm$ 0.19 & 3.47 $\pm$ 0.22 & +228 $\pm$ 13 & --38 $\pm$ 13 & 231 $\pm$ 15 & 274 $\pm$ 18 & 100 $\pm$ 4 \\
	F &	+3.23 $\pm$ 0.57 & --0.57 $\pm$ 0.57 & 3.28 $\pm$ 0.66 & +216 $\pm$ 38 & --38 $\pm$ 38 & 219 $\pm$ 44 & 260 $\pm$ 53 & 100 $\pm$ 10 \\

	\noalign{\vspace{0.5mm}}
	\hline

	\end{tabular}

	\begin{list}{}{}
		\item[$^{\mathrm{a}}$] Position Angle: North 0\,$\degr$, East 90\,$\degr$, South =180\,$\degr$, West 270\,$\degr$.
	\end{list}

\end{table*}

\subsection{Optical shocks}

Our optical deep field (Fig.\,\ref{B335_opt}) shows five HH objects (HH\,119\,A--E) in an almost exact line along constant declination.
This is a very clear case of a bipolar HH flow, with proper motions (Table.\,\ref{propermot_opt}) and bow shock morphologies that would
be expected from their positions relative to the outflow source. The projected space velocities of the optically bright and neatly lined up HH
objects HH\,119\,A--D (200--280\,km\,s$^{-1}$) are much faster than the near-IR bright SSE moving groups 3 and 4 (15--75\,km\,s$^{-1}$).

One explanation for this could be if the fast moving, east-west flow, contains gas expelled into an already cleared out bipolar cone by the outflow
source. The optically bright shocks, HH objects, could then be caused by faster gas moving into slower gas (with shock velocities much lower
than the space velocities measured from proper motions). However, at the edge of the cone-shaped cavity the expelled gas could interact more with
the surrounding globule itself via turbulence, making these near-IR shocks have slower space velocities, almost stationary, due to the encounter with
the probably slow moving gas of the globule.

Given that several HH objects are now known on each side of the outflow source, all having similar space velocities, the shocks within each lobe
have different dynamical ages and there has thus been {\it repeated} eruptive episodes where matter from the accretion is expelled into the outflow 
lobes. Objects B and C are located symmetrically about the outflow source, moving in opposite directions, and were probably expelled in the same
outburst. Using their (optical) proper motions and distances from the central VLA source we find dynamical ages of
about 250 and 265 years for HH\,119\,B and C, respectively.
The proper motions and dynamical ages of objects B anc C was also measured by Reipurth et al. \cite{reipurth} who suggested that this outburst
may have occurred 350 years ago.

There are four more HH objects seen in the optical deep field, HH\,119\,F--I. Objects F and H are located in the eastern lobe of the E--W flow, close
to the HH\,119\,A--E line, while objects G (east) and I (west) seem to belong to the tilted (roughly ESE--WNW) flow. HH\,119\,I is the westernmost
object in this flow, and an optical counterpart of an H$_2$ knot (IR\,8n) in a system of at least 15 knots. Since these HH objects (and object D) are
new detections, their proper motions are unknown, except for object F which is faintly seen in the H$\alpha$ first epoch mosaic. Its appearance, bow
shaped with a clear separation of H$\alpha$ and [SII], and its location relative to the outflow source agrees well with the calculated proper motion.
The other objects are clearly HH objects even though they lack proper motions, based on their extended nature, separation of H$\alpha$ and [SII], shapes
and locations on the outflow axis. As a further confirmation of HH status for these objects, HH\,119\,E, G and H are clearly seen in the shock
sensitive Spitzer filter centred at 4.5\,$\mu$m (Fig.\,\ref{B335_IRAC}) along with most of the other HH objects. The reason for this
is discussed in Section \ref{spitzersec}.

\begin{figure*}
	\centering
	\includegraphics[width=15.0cm]{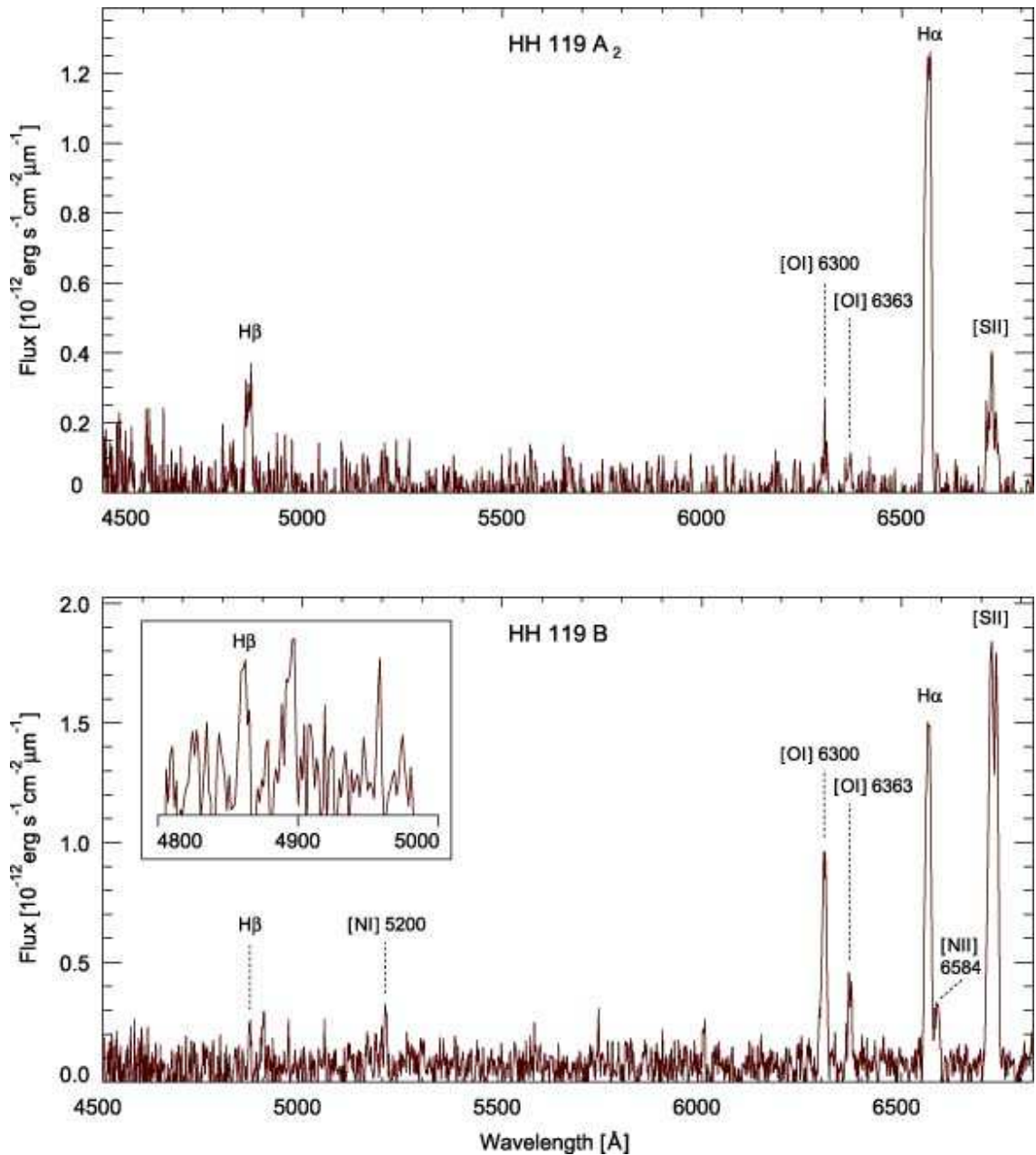}
	\caption{Optical spectra of HH\,119\,A and B (NOT/ALFOSC). Object A is clearly of higher excitation than object B. H$\beta$ is
	detected for the first time in object A and B, as well as strong [OI] lines in HH\,119\,B that were not seen in spectra
	taken 13\,years earlier (suggesting variable [OI] lines).
	}
	\label{HH_spectra1}
\end{figure*}

HH\,119\,E has a different flux distribution than the other HH objects. In Fig.\,\ref{B335_opt} its leading shock is seen more clearly in
the $R$ band than in [SII] emission (making the leading shock look blue in the mosaic even though the $R$ and [SII] position of the shock
overlaps, followed by the H$\alpha$ shock). This is very likely due to unusually bright [OI]~$\lambda\lambda$~6300/63 emission.
The situation is similar to HH\,119\,B, which has bright [OI] emission lines as seen in recent spectra
(Fig.\ref{HH_spectra1}), but with an even higher [OI]/[SII] ratio since HH\,119\,B looks red-yellow-green across the shock, using the same
colour scaling.

HH\,119\,G, seen in both H$\alpha$ and [SII], coincides very well with the brightest peak of the corresponding feature in the
near-IR mosaic (H$_2$ knot IR\,4a) and has a separation between H$\alpha$ and [SII] supporting the near-IR proper motion.

Objects A and E (F) are located roughly symmetrically on opposite sides of the outflow source and may belong to the same outburst, prior to the
B and C outburst. We estimate the ages of HH\,119\,A and F to be about 700 and 590\,years, respectively. The difference of about 100 years in
these estimates may be due to the large uncertainty in proper motion for object F (which is only faintly detected in the first epoch image).


\begin{table*}
	\caption{{\bf Optical spectroscopy.}
	}
	\label{HH_spectable}
	\begin{tabular}{llccccc}
	  \hline
        \noalign{\vspace{0.5mm}}
        Object & Line identification & Flux inside slit$^{\mathrm a}$ & F$^{\mathrm b}$ & F$_{0}$$^{\mathrm c}$ & F($\lambda$6717)/F($\lambda$6731) & n$_e$$^{\mathrm d}$ \\
	       &      & (10$^{-17}$\,erg\,s$^{-1}$\,cm$^{-2}$) & (H$\alpha$) & (H$\alpha$) & & (cm$^{-3}$) \\
 	\noalign{\vspace{0.5mm}}
        \hline

	\noalign{\vspace{1.0mm}}

	HH\,119\,A & H$\beta$ $\lambda$4861 & 47.79 $\pm$ 4.0 & 0.197 $\pm$ 0.020 & 0.304 $\pm$ 0.035 & & \\
		   & [OI] $\lambda$6300       & 20.04 $\pm$ 3.7 & 0.083 $\pm$ 0.017 & 0.087 $\pm$ 0.019 & & \\
		   & [OI] $\lambda$6363	     & 11.73 $\pm$ 3.7 & 0.048 $\pm$ 0.016 & 0.050 $\pm$ 0.017 & & \\
		   & H$\alpha$ $\lambda$6563 & 242.1 $\pm$ 3.6 & 1.000 & 1.000 & & \\
		   & [NII] $\lambda$6583      & 10.45 $\pm$ 2.6 & 0.043 $\pm$ 0.012 & 0.043 $\pm$ 0.012 & & \\
		   & [SII] $\lambda$$\lambda$6717, 6731 & 75.93 $\pm$ 3.7 & 0.314 $\pm$ 0.021 & 0.304 $\pm$ 0.021 & 1.19 (1.04-1.36) & 250 (60-500) \\

	\noalign{\vspace{1.5mm}}

	HH\,119\,B & H$\beta$ $\lambda$4861 & 13.3 $\pm$ 3.3 & 0.056 $\pm$ 0.016 & 0.231 $\pm$ 0.087 & & \\
		   & [NI] $\lambda$$\lambda$5198, 5201 & 14.19 $\pm$ 2.8 & 0.060 $\pm$ 0.013 & 0.167 $\pm$ 0.044 & & \\
		   & [OI] $\lambda$6300	     & 121.9 $\pm$ 4.4 & 0.517 $\pm$ 0.027 & 0.611 $\pm$ 0.033 & & \\
		   & [OI] $\lambda$6363	     & 38.35 $\pm$ 4.3 & 0.163 $\pm$ 0.021 & 0.185 $\pm$ 0.025 & & \\
		   & H$\alpha$ $\lambda$6563   & 235.8 $\pm$ 3.6 & 1.000 & 1.000 & & \\
		   & [NII] $\lambda$6583	     & 26.27 $\pm$ 4.2 & 0.111 $\pm$ 0.020 & 0.110 $\pm$ 0.020 & & \\
		   & [SII] $\lambda$$\lambda$6717, 6731 & 440.4 $\pm$ 4.2 & 1.867 $\pm$ 0.046 & 1.686 $\pm$ 0.043 & 0.82 (0.76-0.90) & 1260 (890-1590) \\

	\noalign{\vspace{1.5mm}}

	HH\,119\,C & [NI] $\lambda$$\lambda$5198, 5201 & 27.64 $\pm$ 4.4 & 0.138 $\pm$ 0.028 & $\sim$0.384 & & \\
		   & [OI] $\lambda$6300	     & 81.46 $\pm$ 4.3 & 0.408 $\pm$ 0.037 & $\sim$0.482 & & \\
		   & [OI] $\lambda$6363	     & 35.07 $\pm$ 3.7 & 0.176 $\pm$ 0.026 & $\sim$0.200 & & \\
		   & H$\alpha$ $\lambda$6563  & 199.6 $\pm$ 7.0 & 1.000 & 1.000 & & \\
		   & [NII] $\lambda$6583	     & 19.47 $\pm$ 5.9 & 0.098 $\pm$ 0.034 & $\sim$0.097 & & \\
		   & [SII] $\lambda$$\lambda$6717, 6731 & 327.3 $\pm$ 5.0 & 1.639 $\pm$ 0.085 & $\sim$1.481 & 1.29 (1.12-1.40) & 140 (30-360) \\

	\noalign{\vspace{1.5mm}}

	HH\,119\,D & [OI] $\lambda$6300       & 16.19 $\pm$ 6.8 & 0.53 +0.44-0.28 & ... & & \\
		   & H$\alpha$ $\lambda$6563 & 30.58 $\pm$ 6.7 & 1.000 & ... & & \\
		   & [SII] $\lambda$$\lambda$6717, 6731 & 104.2 $\pm$ 6.8 & 3.41 +1.24-0.80 & ... & ... & ... \\

	\noalign{\vspace{0.5mm}}
        \hline

	\end{tabular}

	\begin{list}{}{}
		\item[$^{\mathrm{a}}$] Using a slit width of 2\farcs5 along constant declination and centred on HH\,119\,B.
		\item[$^{\mathrm{b}}$] Observed flux, relative to H$\alpha$.
		\item[$^{\mathrm{c}}$] Dereddened flux, using the extinction from the model results found in Table~\ref{physics_HH}.
		\item[$^{\mathrm{d}}$] Electron density. Estimate of uncertainty range given in parenthesis.
	\end{list}

\end{table*}

\begin{figure*}
	\centering
	\includegraphics[width=15.0cm]{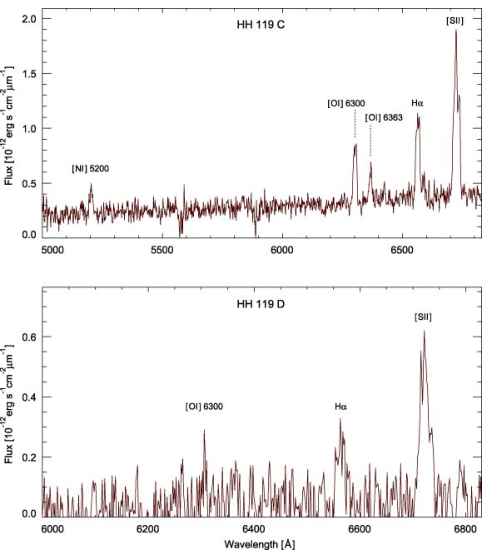}
	\caption{Optical spectra of HH\,119\,C and D (NOT/ALFOSC). These are the first spectra of HH\,119\,C and D, showing
	both to be of low excitation. HH\,119\,C is close to a bright star, therefore some continuum is seen in its spectrum.
	}
	\label{HH_spectra2}
\end{figure*}

\subsection{Optical spectra and line-ratios of HH\,119\,A--D}

The optical spectra of HH\,119\,A--D are shown in Figures\,\ref{HH_spectra1} and \ref{HH_spectra2}. Even though all spectra cover the spectral
range 3850--6850\,\AA, each spectrum has been plotted using individual ranges, omitting parts where no features are seen.

In Table~\ref{HH_spectable} we present calibrated and relative (to H$\alpha$) line fluxes for all detected lines. The dereddened flux (F$_0$)
has been calculated using the extinctions we find for HH\,119\,A and B using planar shock models and comparing the intrinsic and
observed H$\alpha$/H$\beta$ ratios (cf Sect.~\ref{shockmodels}). We have also calculated the mean electron density, n$_e$, in HH\,119\,A, B
and C for the position covered by the slit using the [SII] I($\lambda$6717)/I($\lambda$6731) line ratio. For this we use the effective
collision strengths for electron impact excitation given in Ramsbottom et al.~\cite{ramsbottom} and Einstein A-coefficients from
Keenan et al.~\cite{keenan}, assuming an electron preshock temperature of $\sim$\,10$^{4}$\,K which is a common assumption for HH
shock models (see e.g. Morse et al.~\cite{morse94}). The uncertainty range in these estimates are given in paranthesis in
Table~\ref{HH_spectable}.

All objects show H$\alpha$ in strong emission and except for HH\,119\,A, the [SII]\,6717/31 doublet is the strongest line.
The choice of nb filters for our optical deep fields were in fact based on the line strengths in these spectra.
HH\,119 A, B and C also have [OI]~$\lambda\lambda$~6300/63 emission lines ([OI]\,6300 faintly seen in object D as well but with poor S/N).

It is also very interesting to note that the [OI] lines are strong in HH\,119\,B. This is somewhat unexpected, as they were not detected
in previous spectra by Reipurth et al.\cite{reipurth} taken 13 years earlier even though the S/N of those observations should have been more
than enough for a clear detection.
This suggests that the [OI] lines are very variable and that conditions have changed a lot in HH\,119\,B during this time period.
The spectra confirm that object A is an intermediate excitation object while HH\,119\,B, C and D are found to have lower excitation.
Since HH\,119\,A is located further out from the outflow source than object B, the lower excitation of B could be explained
by it moving in the wake of the bow shock from object A. 

It is common practice to plot contours of e.g. H$\alpha$--[SII] to illustrate regions dominated by H$\alpha$ (positive) and [SII] (negative), 
respectively. This has the disadvantage of cancelling out in regions where H$\alpha$ and [SII] have similar flux. It also lowers the signal of
both the H$\alpha$ and [SII] contours while the noise remains at the same level. This is the case since they always cancel out partially as long
as there is some H$\alpha$ or [SII] flux present. A map of F([SII])/F(H$\alpha$) is better in that respect but has its own disadvantages, the most
noticeable one being that very large values (both positive and negative) are found where F(H$\alpha$) is close to zero (mostly in sky regions).
It is also common to plot H$\alpha$+[SII] contours. This uses the full S/N for contours but hides the relative flux information
that indicates the abundance of the two species.

\begin{figure*}
	\centering
	\includegraphics[height=17cm]{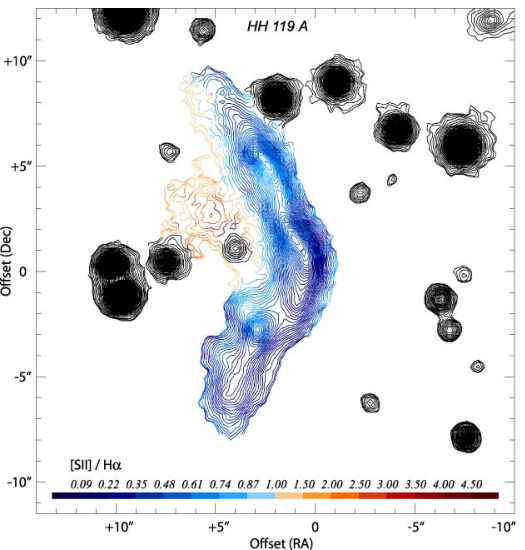}
	\caption{
	Detailed overview of HH~119\,A simultaneously showing a H$\alpha$+[SII] surface brightness map (contours) and the excitation
	structure by colouring the contours based on the [SII]/H$\alpha$ ratio.
	Note that the colour coding is totally independent of the contours, this means that any given contour
	can have any variation of colours ([SII]/H$\alpha$) along its curve. Stars are also included in this plot, these
	however have black contours.
	HH~119\,A is centred so that its brightest position is roughly located at zero offset. 
	The faintest H$\alpha$+[SII] contour shown has a surface brightness of
	about 8.4\,$\times$\,10$^{-17}$\,erg\,s$^{-1}$\,cm$^{-2}$\,arcsec$^{-2}$.
	}
	\label{HH119A_rat}
\end{figure*}

\begin{figure*}
	\centering
	\includegraphics[height=17cm]{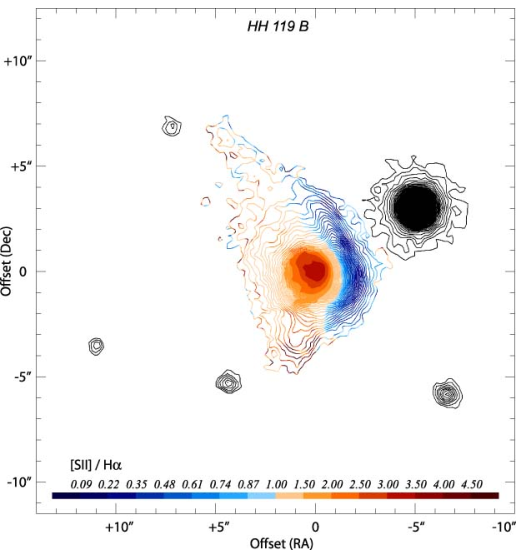}
	\caption{
	Detailed overview of HH~119\,B, using the same technique as in Figure~\ref{HH119A_rat}. HH~119\,B is centred so that its brightest
	position is roughly located at zero offset. The faintest H$\alpha$+[SII] contour shown has a surface brightness of about
	3.1\,$\times$\,10$^{-17}$\,erg\,s$^{-1}$\,cm$^{-2}$\,arcsec$^{-2}$.
	}
	\label{HH119B_rat}
\end{figure*}

A solution to this illustrative dilemma is found by e.g. using coloured contours to simultaneously plot both the total
flux F(H$\alpha$)+F([SII]) (contours) and their relative strength F(H$\alpha$)/F([SII]) (colours).
This is presented for HH\,119\,A and B in Figures \ref{HH119A_rat} and \ref{HH119B_rat}, respectively.
Note that the contours are coloured on a pixel-scale level (0\,\farcs188) meaning that each contour can have any colours along
its curve. By choosing reasonable surface brightness limits for the faintest contour included in the plots only relevant regions are
included (i.e. most of the HH object and all stars, while excluding the sky).
Technically this is done by multiplying a high-resolution binary contour map (zero for sky and one for contours) pixel-by-pixel
with an equally high-resolution F(H$\alpha$)/F([SII]) ratio map that has been coloured with red and blue shades to separate
H$\alpha$ (blue) and [SII] (red) bright regions.

From these Figures it is clear that HH\,119\,A is of relatively high excitation compared to HH\,119\,B which is a low excitation object.
There is also a clear separation in both HH\,119 A and B, between H$\alpha$ (leading) and [SII] (trailing) in the direction
of motion. In both cases the bow shock is of high excitation (low [SII]/H$\alpha$ ratio) and the emission behind the bow shock is
of low excitation (high [SII]/H$\alpha$ ratio). This structure is typical for most HH objects, H$\alpha$ tracing sharp leading edges
of the cooling zone and [SII] dominating in the chaotic and clumpy postshock region. It is indicative of a heavy flow that rams into
a less dense ambient medium. The situation is the same for all other HH objects in B335, except for E and G which have [SII] leading
H$\alpha$, suggesting a flow that rams into a denser ambient medium. See section \ref{HH119shocks} for more discussion of the individual HH objects.


\begin{table*}
	\caption{{\bf Extinction and shock model results for HH\,119\,A, B and C.}
	}
	\label{physics_HH}
	\begin{tabular}{lccccccccc}
	\hline
        \noalign{\vspace{0.5mm}}
        Iteration step & (H$\alpha$/H$\beta$)$_0$ & A$_V$$^{\mathrm a}$ & E(B-V) & n$_0$$^{\mathrm b}$ & C\,$^{\mathrm c}$ & Ionization I\,$^{\mathrm d}$ & $<I>$\,$^{\mathrm e}$ & Shock velocity V$_S$$^{\mathrm d}$ & $<V_S>$ \\
              & & (mag) & (mag) & (cm$^{-3}$) & & (\%) & (\%) & (km/s) & (km/s) \\
	\noalign{\vspace{0.5mm}}
	(1) & (2) & (3) & (4) & (5) & (6) & (7) & (8) & (9) & (10) \\ 
 	\noalign{\vspace{0.5mm}}
        \hline

        \noalign{\vspace{0.5mm}}
	\multicolumn{10}{c}{HH 119\,A} \\
        \noalign{\vspace{0.5mm}}
	\hline 
	\noalign{\vspace{1.0mm}}

	Initial guess\,$^{\mathrm f}$ & 3.00 & 1.65 & 0.53 & 625 & 20 & ... & 2.0 & ... & 30.0 \\

	Iteration & 3.34 & 1.31 & 0.42 & 25  & 40 & 40.0, 23.0, 13.5 & 25.5 $\pm$ 14 & 74.7, 60.2, 46.4 & 60.4 $\pm$ 15 \\

	{\bf Result} & {\bf 3.29} & {\bf 1.36} & {\bf 0.44} & {\bf 23}  & {\bf 40} & {\bf 40.0, 30.0, 13.5} & {\bf 27.8 $\pm$ 14} & {\bf 74.7, 68.7, 45.3} & {\bf 62.9 $\pm$ 16} \\

	\noalign{\vspace{1.0mm}}
        \hline

        \noalign{\vspace{0.5mm}}
	\multicolumn{10}{c}{HH 119\,B} \\
        \noalign{\vspace{0.5mm}}
	\hline 
	\noalign{\vspace{1.0mm}}

	Initial guess\,$^{\mathrm f}$ & 3.00 & 5.59 & 1.80 & 8400 & 15 & ... & 1.0 & ... & 30.0 \\

	Iteration & 4.33 & 4.44 & 1.43 & 1530 & 25 & 3.1, 2.2, 4.6 & 3.3 $\pm$ 1.3 & 37.8, 34.6, 39.8 & 37.4 $\pm$ 2.7 \\

	{\bf Result} & {\bf 4.33} & {\bf 4.44} & {\bf 1.43} & {\bf 1070} & {\bf 28} & {\bf 3.7, 3.6, 5.2} & {\bf 4.2 $\pm$ 1.0} & {\bf 37.1, 35.7, 38.0} & {\bf 36.9 $\pm$ 1.5} \\

	\noalign{\vspace{1.0mm}}
        \hline

        \noalign{\vspace{0.5mm}}
	\multicolumn{10}{c}{HH 119\,C} \\
        \noalign{\vspace{0.5mm}}
	\hline 
	\noalign{\vspace{1.0mm}}

	Initial guess\,$^{\mathrm f}$ & 4.33 & (4.44)\,$^{\mathrm g}$ & (1.43)\,$^{\mathrm g}$ & 930 & 15 & ... & 1.0 & ... & 30.0 \\

	Iteration & 4.18 & (4.44)\,$^{\mathrm g}$ & (1.43)\,$^{\mathrm g}$ & 100 & 28 & 6.1, 4.1, 6.1, 3.3 & 4.9 $\pm$ 1.5 & 43.6, 37.7, 38.1, 33.5 & 38.2 $\pm$ 4.2 \\

	{\bf Result} & {\bf 4.38} & (4.44)\,$^{\mathrm g}$ & (1.43)\,$^{\mathrm g}$ & {\bf 83} & {\bf 23} & {\bf 8.6, 7.6, 7.7, 5.5} & {\bf 7.3 $\pm$ 1.5} & {\bf 41.9, 37.2, 33.3, 26.9} & {\bf 34.8 $\pm$ 6.4} \\

	\noalign{\vspace{1.0mm}}
        \hline

	\end{tabular}

	\begin{list}{}{}
		\item[$^{\mathrm{a}}$] Using the extinction law of Cardelli et al. \cite{cardelli}, R$_V$ = 3.1 and (H$\alpha$/H$\beta$)$_{obs}$.
		\item[$^{\mathrm{b}}$] Preshock density, $n_0 = (n_e/<I>)/C$.
		\item[$^{\mathrm{c}}$] Shock compression.
		\item[$^{\mathrm{d}}$] Estimates from line ratios [OI]\,6300/H$\alpha$, [SII]\,6717+31/H$\alpha$, [NII]\,6583/[OI]\,6300. For HH\,119\,C also from [NI]\,5200/[NII]\,6583. 
		\item[$^{\mathrm{e}}$] [SII]-weighted ionization fraction $<I>$.
		\item[$^{\mathrm{f}}$] Initial guesses taken from objects with similar excitation, HH\,47\,jet (HH\,119\,A) and HH\,34\,jet (B and C). See Hartigan et al.\cite{hartigan94}.
		\item[$^{\mathrm{g}}$] Since H$\beta$ is not detected from HH\,119\,C, the extinction is $\it assumed$ to be the same as for HH\,119\,B (see text).
	\end{list}

\end{table*}

\subsection{Shock modeling and extinction}
\label{shockmodels}

In this section we use the planar shock models of Hartigan et al.~(\cite{hartigan94}) with our electron densities and different line ratios
(from Table~\ref{HH_spectable}) for HH\,119\,A--C to calculate the average shock velocity ($<V_S>$), [SII]-weighted ionization
fraction ($<I>$), preshock density (n$_0$), compression (C) and intrinsic (H$\alpha$/H$\beta$)$_0$ ratio of each shock.
Using the observed (H$\alpha$/H$\beta$) ratio we also estimate the extinction to HH\,119\,A and B by applying the optical extinction law described
in Cardelli et al.~(\cite{cardelli}) for $R_V = 3.1$, and calculating the extinction A$_V$ needed to match the intrinsic ratio from the model.

This is an iterative process since we need to know some of the model results, the preshock density n$_0$ (for the shock models) and the
(H$\alpha$/H$\beta$)$_0$ ratio (to deredden the line ratios) to apply the models to our observations. The preshock density can be found
from the models if we can estimate the compression, ionization $<I>$ and shock velocity V$_S$ to apply to our electron density since
$n_0 = (n_e/<I>)/C$. The intrinsic (H$\alpha$/H$\beta$) ratio can be assumed to be $\sim$3 as a first approximation (case B recombination, but
usually higher than this for HH objects). By making a few iterations through the shock models and recalculating the extinction and line ratios
each time we can get a converging solution. However, for each object we need initial rough estimates of the shock parameters. For this we compare
our spectra with the examples given in Hartigan et al.~\cite{hartigan94}.

HH\,119\,A is an intermediate excitation object, reminiscent of the HH\,47 jet. We therefore start by assuming that
V$_S$\,$\sim$~30\,km\,s$^{-1}$, $<I>$\,$\sim$\,2\,\% and C\,$\sim$\,20 (from V$_S$ and the shock models). Using our observationally determined
n$_e$\,$\sim$\,250\,cm$^{-3}$ we estimate that n$_0$\,$\sim$\,625\,cm$^{-3}$. For a (case B) (H$\alpha$/H$\beta$)$_0$\,$\sim$\,3 and the observed
value 5.076 the extinction becomes $A_V = 1.65$ and $E(B-V) = 0.53$. The dereddened line ratios [OI]~6300/H$\alpha$, [SII]~6717+31/H$\alpha$ and
[NII]~6583/[OI]~6300 then gives three estimates of V$_S$, $<I>$, C and (H$\alpha$/H$\beta$)$_0$, one for each line ratio. These values are then
used for the next iteration.

HH\,119\,B and C are of low excitation, similar to the HH\,34 jet (Hartigan et al.~\cite{hartigan94}), we therefore make the initial estimates
V$_S$\,$\sim$~30\,km\,s$^{-1}$, $<I>$\,$\sim$\,1\,\% and C\,$\sim$\,15.

The fact that we do not detect [OIII]\,$\lambda$5007 in any of the objects suggests (Morse et al.~\cite{morse93}) that the shock velocities
involved are {\it lower} than 90\,km\,s$^{-1}$ (for no preionization), 80\,km\,s$^{-1}$ (equilibrium preionization) or
40\,km\,s$^{-1}$ (fully preionized). This, in addition to the excitation types, supports our initial estimates of slow shock speeds
(V$_S$\,$\sim$\,30\,km\,s$^{-1}$).

The initial estimates, iterations and results are given in Table~\ref{physics_HH}. Since we do not detect H$\beta$ in the spectrum of HH\,119\,C
its extinction is assumed to be roughly the same as for object B, motivated by the fact that their projected distances from the outflow source
are about the same. The resulting extinctions are found to be $A_V = 1.36$ (1.05-1.67), $E(B-V) = 0.44$ (0.34 - 0.54) for HH\,119\,A and
$A_V = 4.44$ (3.67-5.28), $E(B-V) = 1.43$ (1.18 - 1.70) for HH\,119\,B.
The assumption that object C has the same extinction as object B based on equal distance from the outflow source may however be inaccurate, as
B335 has been shown to have a long tail (Frerking et al. \cite{frerking87}), a remanence from its formation history as a cometary
globule (Reipurth \cite{reipurth83}).
However, since we only use line ratios close to each other in wavelength (a few hundred \AA), except for the [NI]~5200/[NII]~6583 ratio, the
model results for HH\,119\,C should not be very sensitive to this assumption.

It is interesting to note that shock velocity and ionization found for object A is much higher than that for object B. This agrees with the
suggestion that object B is following in the wake of the outermost object A and thus colliding with cloud material (or turbulence in the flow
itself) that has already been sped up by object A. The higher excitation of HH\,119\,A also agrees with a higher shock velocity and ionization.
HH\,119\,B is found to have much higher preshock density n$_0$ than object A, which might be expected given its location much closer to the centre
of the globule where the density is higher.

\begin{figure*}
	\centering
	\includegraphics[width=18cm]{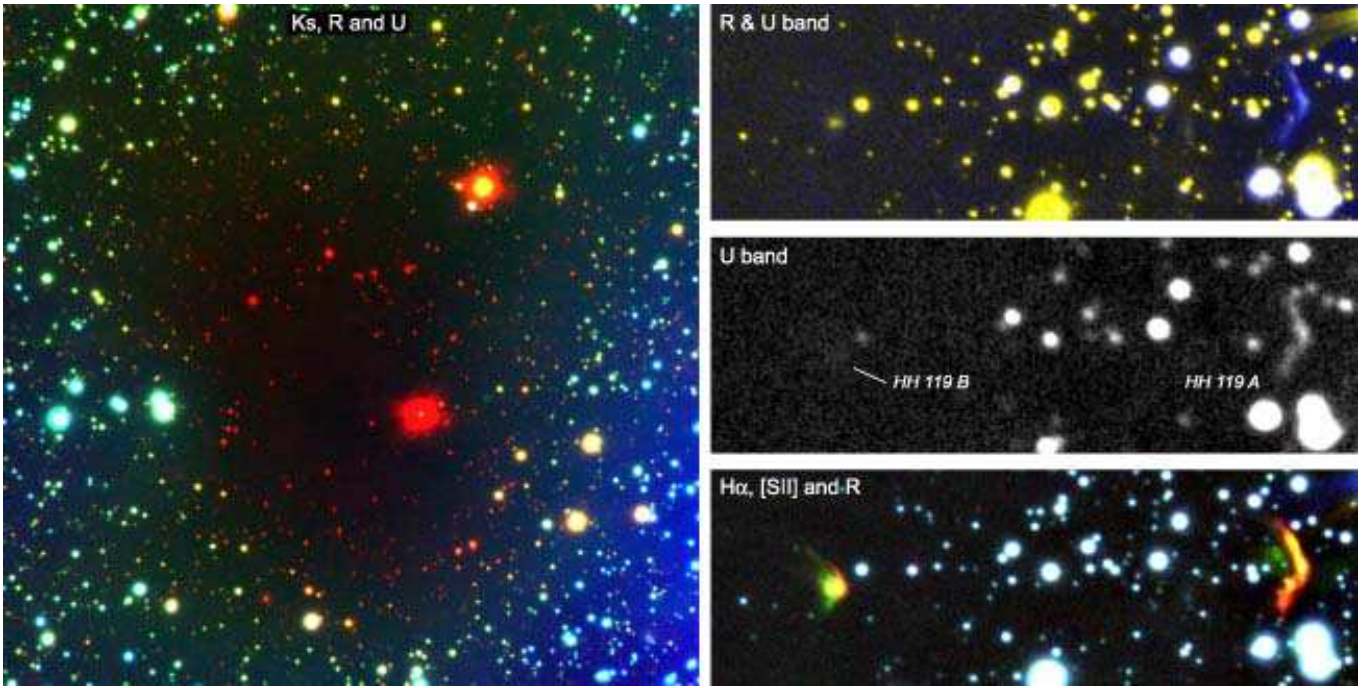}
	\caption{Near-UV deep field. A composite of $K_S$ (red), $R$ (green) and $U$ (blue) mosaics (left panel). The cloud extinction is effectively
	shown using this filter combination, spanning from the near-UV to the near-IR. After removing the background from scattered starlight in the
	$U$ mosaic, three of the HH objects are detected (HH\,119\,A, B and E). In the right panel we show several filter combinations
	using background subtracted mosaics. It is clear that the appearance of object A is very similar in the $U$ and H$\alpha$ filters.
	The field of view is 4\farcm25\,$\times$\,4\farcm13 in the left panel and 116$\arcsec$\,$\times$\,39$\arcsec$ in each of the
	right panels.
	}
	\label{HH119_U}
\end{figure*}

\subsection{Near-UV imaging}

In Figure~\ref{HH119_U} our $U$ band deep field is presented. HH\,119\,A is clearly detected, and has a very similar appearance to
its H$\alpha$ counterpart. After background subtraction of the scattered starlight in the globule we also detect HH\,119\,B and E but
these are however much fainter than object A in the near-UV. For object B this is expected because of its much higher extinction
($\Delta$A$_V$\,$\approx$\,3). None of the other HH objects are detected, but for HH\,119\,C and D this could partly be explained by
their proximity to a bright star combined with the seeing in the near-UV.

The transmission curve of the $U$ filter is shown in Fig.~\ref{Ufilter} together with the atmospheric transmission for the mean airmass of the
observations, the detector QE and the total system transmission (black solid curve). There are probably only two spectral features that contribute
to the $U$ band flux of these HH objects. The [OII]\,$\lambda$3728 line, located close to the maximum system transmission, which has also been shown
to be strong in other HH objects (see e.g. Hartigan et al. \cite{hartigan99}) and the blue continuum from two-photon emission. The blue
continuum spectral distribution has been plotted in Fig.~\ref{Ufilter} using a blue dashed curve with a scale corresponding to the percentage
of its peak value. The total $U$ band flux for the three detected HH objects are given in Table~\ref{HH_fluxtable}.

\subsection{Spitzer mid-IR imaging}

\label{spitzersec}
\begin{figure*}
	\centering
	\includegraphics[width=18cm]{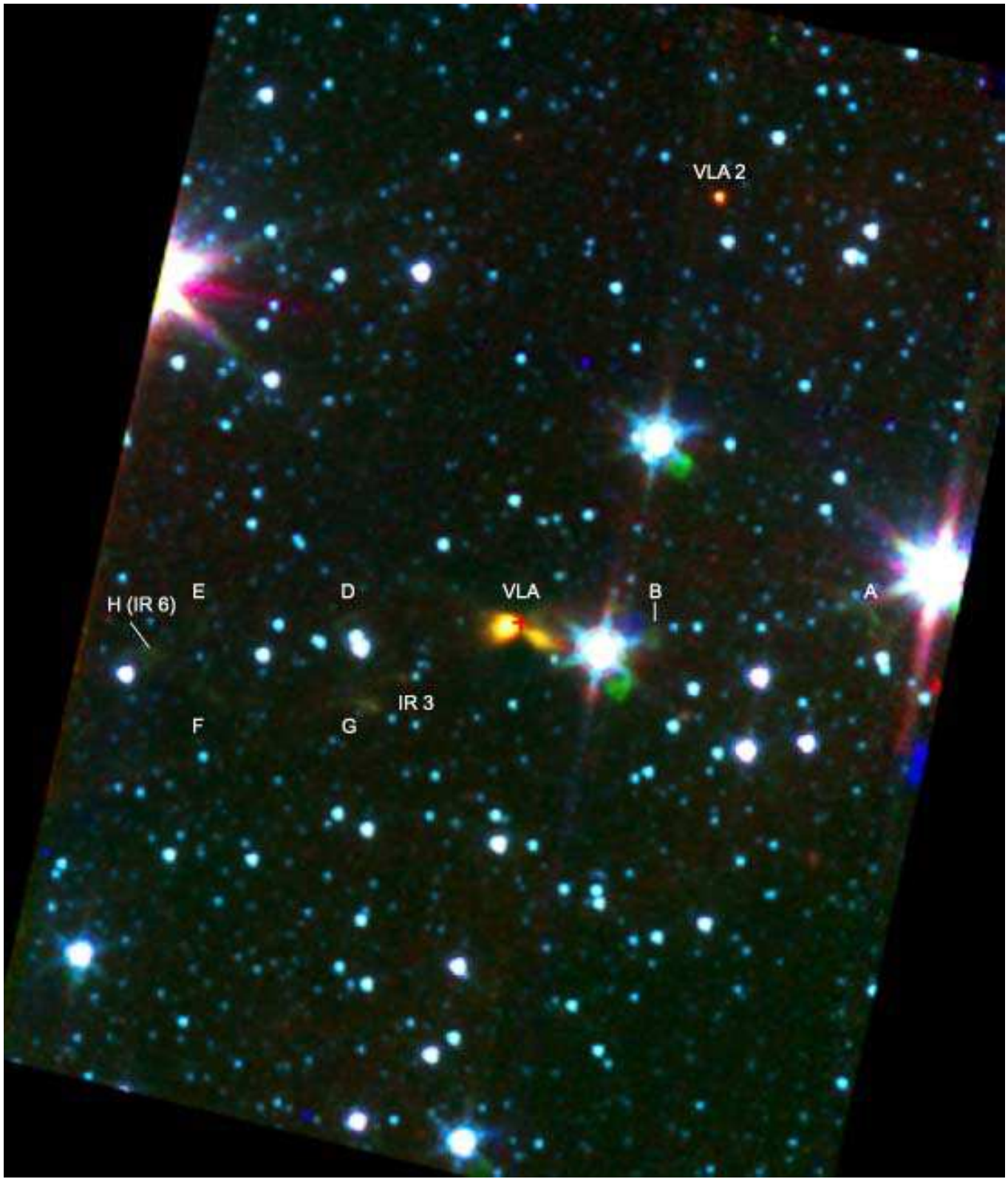}
	\caption{Spitzer IRAC composite of B335 using a 3.5\,$\mu$m (blue), 4.5\,$\mu$m (green) and 8.0\,$\mu$m (red) colour coding. The red
	plus sign marks the central VLA source position. There is a clear correlation between the bipolar cone-shaped structure seen at these
	wavelengths (3.5--8.0\,$\mu$m) and the VLA source position. Many of the HH objects and S(1) knots seen in our optical and near-IR
	mosaics are also detected at 4.5\,$\mu$m (IRAC channel 2) which is clearly the most efficient channel for tracing the flows.
	}
	\label{B335_IRAC}
\end{figure*}

Figure\,\ref{B335_IRAC} shows our Spitzer 3.5, 4.5 and 8.0\,$\mu$m IRAC composite of B335.
An hourglass-shaped reflection nebula is clearly seen in the centre (at 4.5 and 8.0\,$\mu$m), with an outflow cone to each side of
the central VLA source (IRAS\,19345+0727), marked by a cross in the Figure. Most of the HH objects we see in our optical and
near-IR deep fields can also be seen in the 4.5\,$\mu$m channel of Spitzer (green in the colour composition used in Fig.\,\ref{B335_IRAC}).
Note that, except for the strong artefacts seen close to bright stars (to the SSW) in this filter, all these extended objects are most
likely HH objects. These include all HH objects A--H, except HH\,119\,C, which is too close to a bright star for detection. HH\,119\,I is located
outside the Spitzer image.

It has been shown in several other Spitzer surveys (e.g. Noriega-Crespo et al. \cite{noriega04}, Harvey et al. \cite{harvey06} and
G\aa lfalk \& Olofsson \cite{galfalk06}) that the IRAC channel centred at 4.5\,$\mu$m is very efficient in detecting bona-fide Herbig-Haro objects.
The reason for this is partly that the spectral response function is highest in this channel, but there are at least two more contributing factors.
Between approximately 4--5\,$\mu$m there are many vibrational and rotational H$_2$ emission lines, these have been modeled
by Smith \& Rosen (\cite{smith}) for all IRAC bands using three-dimensional hydrodynamic simulations of molecular jets. The strongest
integrated H$_2$ emission is predicted to arise from band 2 because of rotational transitions. For the typical conditions of low-mass
outflows, pure-rotational transitions like S(11)--S(4) (4.18--8.02\,$\mu$m) can actually be much brighter than the
standard 2.12\,$\mu$m H$_2$ line (Kaufman \& Neufeld \cite{kaufman}). Channel 2 is also the most ``PAH-free'' band of IRAC, greatly enhancing its
usefulness as a HH tracer, as opposed to the 5.8 and 8.0\,$\mu$m channels which have very high background contrasts
caused by extended Polycyclic Aromatic Hydrocarbon (PAH) emission, hiding the shock-excited H$_2$ features of the HH flows.

A closer inspection of our IRAC composite actually reveals one more extended object with strong 4.5\,$\mu$m flux, this object
is located to the east of HH\,119\,E and F, at a declination that is consistent with it being part of the HH flow.
It is located just outside the eastern border of our 2.12\,$\mu$m mosaic but a faint object is hinted at this position in
our H$\alpha$ mosaic, although very close to the border where the S/N is degraded due to the few number of overlapping exposures used in that
part of the mosaic. Given its location and brightness in the Spitzer image, together with its H$\alpha$ detection (although faint), we classify
this as an HH object (HH\,119\,H).

We identify a very red object to the north in Fig.\,\ref{B335_IRAC} as HH\,119\,VLA\,2 (Avila et al. \cite{avila}), also known as
Barn\,335\,3 (Anglada et al. \cite{anglada}). The nature of this source is unknown. It is located outside the northern border of our
2.12\,$\mu$m H$_2$ S(1) mosaic and no H$\alpha$ emission is seen near this object. 

\subsection{Individual Herbig-Haro objects}
\label{HH119shocks}

Figure\,\ref{Shocks_optical} shows all the HH objects in detail with individually adapted contrast. In this Section details
of each HH object will be discussed, referring to Fig.\,\ref{Shocks_optical} in the following if the image of a shock is discussed without
further reference.

\begin{figure*}
	\centering
	\includegraphics[width=18cm]{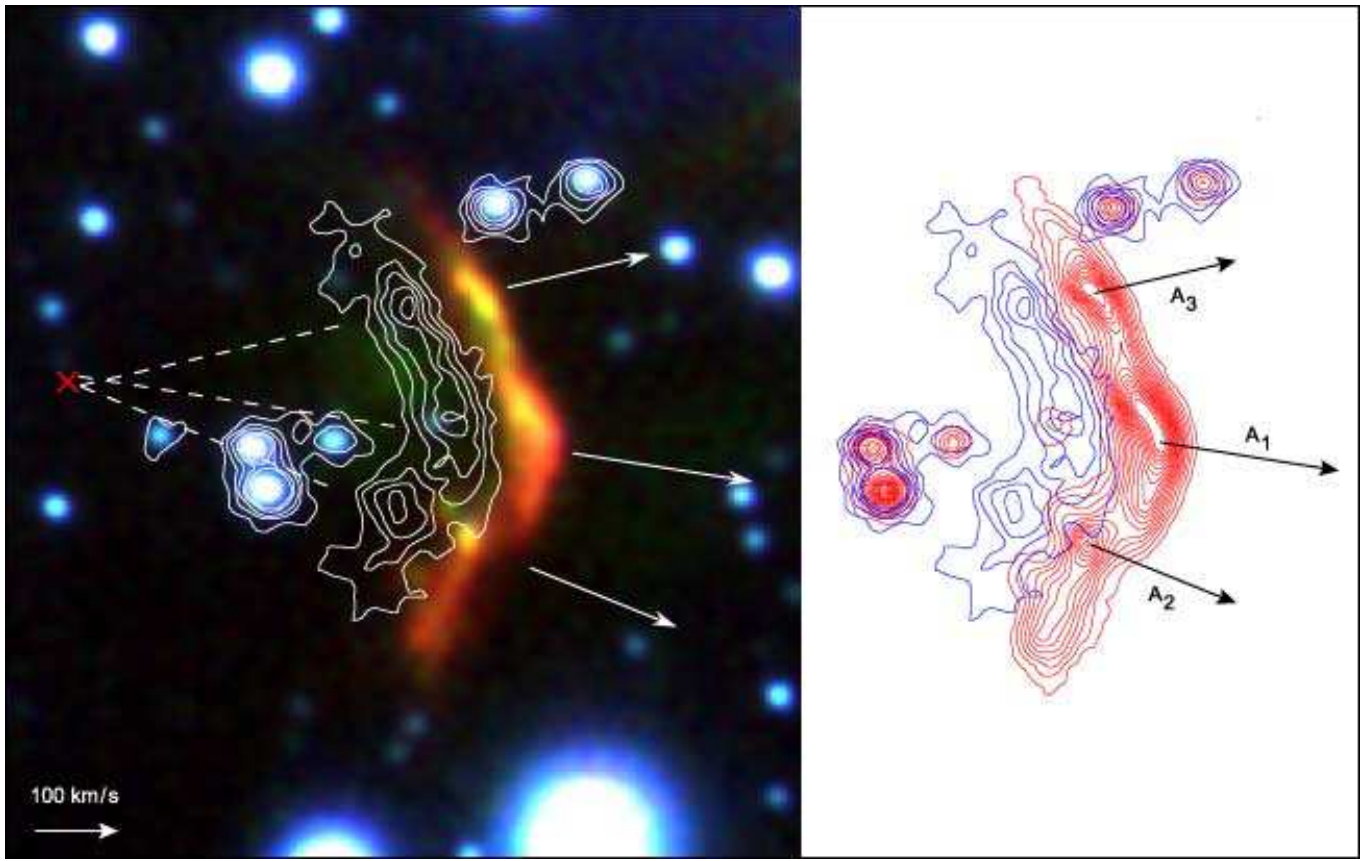}
	\caption{Optical close-up of the HH119\,A shock in H$\alpha$ (red), S[II] (green) and $R$ (blue). The overplotted H$\alpha$ contours from
	the first epoch (15 years earlier, Reipurth et al. \cite{reipurth}) clearly illustrates the movement of this shock.
	Proper motions have been calculated for three H$\alpha$ peaks in the shock (A$_1$, A$_2$ and A$_3$) as shown by the arrows. The shock is clearly
	expanding much faster than would be suggested by its position in relation to the outflow source if a linear expansion was assumed. The
	field size in the left panel is 25\farcs9\,$\times$\,27\farcs8.
	}
	\label{HH119A}
\end{figure*}

\begin{figure*}
	\centering
	\includegraphics[width=18cm]{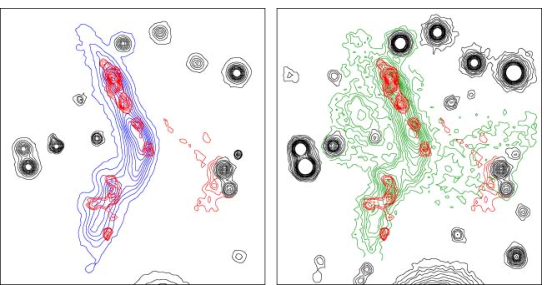}
	\caption{HH\,119\,A flux contours in H$\alpha$ (blue), S[II] (green) and S(1) (red).
	The contour levels have been chosen to emphasize the different features seen in each transition.
	It is clear that most S(1) knots in HH\,119\,A lie well within the optical bow shock. In S[II] there are two more features when
	compared to H$\alpha$, a bright trailing structure and an extended faint feature in front of the bow. This
	feature can also be seen faintly in S(1) as a bow-shaped object leading the main optical bow shock. Stars have been coloured
	black. The field size is 21$\arcsec$\,$\times$\,21$\arcsec$ in both panels.
	}
	\label{HH119A_contours}
\end{figure*}

\subsubsection{HH119 A and IR\,7}
\label{HH119Asec}
The image of HH\,119\,A has been deconvolved using maximum entropy to bring out more details, made possible by the high S/N ratio achieved
for this object. The leading bow shock is dominated by H$\alpha$ emission (even though there is considerable [SII] emission as well)
followed by a region dominated by [SII] emission. There are three condensations (emission peaks) in the bow
shock, called $A_1$ (middle peak), $A_2$ (south) and $A_3$ (north).

We have measured the proper motions within the bow shock using the three H$\alpha$ peaks. The result is shown in Fig.\,\ref{HH119A} with
projected velocities illustrated by arrows. The image is an extreme close-up of the HH\,119\,A part of our deep field, with overplotted
H$\alpha$ contours adapted from Fig.\,2 of Reipurth et al. \cite{reipurth}.
Note that even though the northern condensation ($A_3$) may seem to move in a slightly different direction (not as far north) than the arrow
implies in the left panel of Fig.\,\ref{HH119A}, this is just a visual effect caused by both H$\alpha$ and S[II] being shown as red and
green, respectively, in the colour composite. The proper motions were calculated from the locations of the H$\alpha$ emission knots, as shown in the
right panel.

We note that the central peak found in the first epoch image coincides with a star behind
the shock that is now revealed 15 years later. The H$\alpha$ contribution from this star in the first epoch is unknown, but a wider peak has been
used to trace the proper motion of the middle part of the shock in order to minimize the risk of confusion with this star.

The bow shock is clearly expanding at a much faster rate than would be expected by its distance from the outflow source assuming a linear
expansion, since tracing the arrows backwards leads to a point much further west than the outflow source (Figures\,\ref{B335_opt} and \ref{HH119A}).
The expansion rate has thus increased with time as the shock gradually moves into less dense parts of the globule, approaching
free expansion. Object B has a similar appearance to object A, albeit on a smaller scale, suggesting the amount of expansion between these
positions in the flow. The situation is reminiscent of the HH34\,jet and bow shock (Figures\,9 and 10 of Reipurth et al.~\cite{reipurth02b}) where
the width of the jet and the size of HH objects increase with distance along the jet from the outflow source.

The spectrum of HH\,119\,A$_2$ (Fig.\,\ref{HH_spectra1}) shows strong H$\alpha$ and weak [SII] emission, HH\,119\,A is thus a high excitation object as
opposed to the other HH objects for which we have optical spectra. The first detection of H$\beta$ in object A is also seen in this spectrum.

The picture of HH\,119\,A as a coherent bow shock (as seen in H$\alpha$) is not always true. We detect the S(1) counterpart of object A as broken
up into three parts (IR\,7a--c, see Fig.\,\ref{New_S1_flow}).
In Figure\,\ref{HH119A_contours} we plot flux contours in H$\alpha$, [SII] and the S(1) line. It is evident that most of the
S(1) knots in the northern and southern part (IR\,7a and b) lie well within the contours of the optical bow shock. There is also a third, very
faint counterpart in S(1) (IR\,7c), also bow-shaped, which preceeds the optical bow. Comparing the [SII] and S(1) contours (right panel) we see
that, not only is there extended [SII] emission in a cooling zone behind the bow shock, there is also a faint [SII] structure preceeding the bow, at
the position of IR\,7c.

\begin{figure*}
	\centering
	\includegraphics[width=18cm]{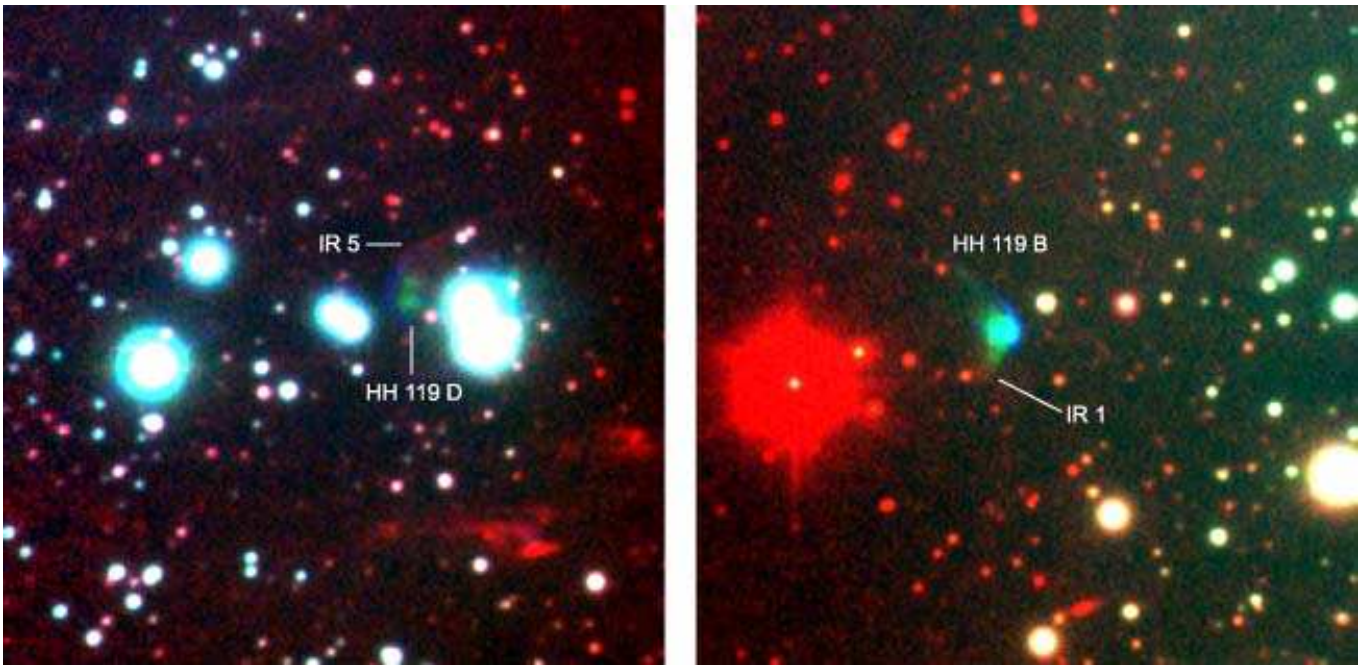}
	\caption{Optical and Near-IR composites of HH\,119\,D\,/\,IR\,5 (left) and B\,/\,IR\,1 (right) using a
	H$\alpha$ (blue), [SII] (green) and 2.12\,$\mu$m H$_2$ (red) colour coding. In both shocks the optical shock emission is clearly strongest
	close to the apex of the bow shock, H$\alpha$ at the front and [SII] in the cooling zone behind it. The near-IR shock
	emission is displaced to the north and south in HH\,119\,D and B, respectively. In the case of HH\,119\,D, the
	near-IR shock is also bow-shaped, but located further to the north. The near-IR counterpart of HH\,119\,B, IR\,1, does not have
	the appearance of a bow, it does however coincide with the far southern wing of the optical bow shock, as might be expected since the
	projected shock velocity (perpendicular to the bow surface) is much lower far out in the wing than at the apex.
	Both fields have a size of 75$\arcsec$\,$\times$\,75$\arcsec$.
	}
	\label{B335_IRopt}
\end{figure*}

\subsubsection{HH119 B and IR\,1}
\label{HH119Bsec}
HH\,119\,B is a nice well-defined bow shock with H$\alpha$ followed by [SII] emission, and an apex with both strong H$\alpha$
and [SII] emission. The shock follows in the wake of HH\,119\,A, which is a possible explanation why the spectrum of this shock shows a
lower excitation (see Fig.\,\ref{HH_spectra1}).
The spectrum also shows object B to have strong [OI]~$\lambda\lambda$~6300/63 emission lines, which was not the case in the Aug. 1990 observations
of Reipurth et al. (\cite{reipurth}). The conditions in object B have apparently changed during the 13\,years between the two epochs.

The near-IR H$_2$ S(1) source IR\,1 has a similar position and proper motion position angle as object B, although their proper motions differ
by more than 70\,km\,s$^{-1}$. To clarify the situation a colour composite using both the optical and near-IR shock observations has been
made (Fig.\,\ref{B335_IRopt}, right panel) by warping the S(1) image to match the pixel size and distortion of the optical
images (H$\alpha$ and S[II]).

These shocks are likely physically connected, with the S(1) bright shock IR\,1 located in the southern part of optically
bright shock B. While their difference in projected velocities suggest that their alignment could be merely
coincidental, the most likely situation is that they are in fact counterparts of the same object, since IR\,1 is located far out in the wing of bow
shock B where the (projected) shock speed is much slower than at the apex. This could explain the S(1) brightness if H$_2$ is not disassociated in
the bow wing. The geometry and expected excitation structure (Fig.\,\ref{B335_IRopt}) thus suggest that IR\,1 is the near-IR counterpart of HH\,119\,B.

\subsubsection{HH119 C, D and IR\,5}
In the image of object C in Fig\,\ref{Shocks_optical}, a bright star has been removed at the position marked by a plus sign in all three
filters (H$\alpha$, [SII] and $R$) using the PSF of a similarly bright star, lowering the contrast considerably and making it possible to
display HH\,119\,C in much more detail. A star, not seen prior to the PSF removal, is also revealed very close to the north of the removed
bright star (at the edge of the black circle).

This is the first detection of object D in the optical, it had only been seen faintly in the near IR prior to this paper (IR\,5). Our near-IR
observations show this to be a bow shock at 2.12\,$\mu$m H$_2$ as well. Object D has a complicated excitation structure, as suggested by the
optical/IR composite image (Fig.\ref{B335_IRopt}). A leading bow-shock dominated by H$\alpha$ is clearly seen, followed by two
well-defined 'sulphur bullets' in the complicated cooling zone and a near-IR H$_2$ bow shock, displaced to the north relative to the optical bow shock.
The situation appears to be similar to that of objects B and IR\,1, with a S(1) strong part of the bow shock far out in the (north) wing where the
projected shock speed is much lower than at the apex of the bow shock.

\subsubsection{HH119 E, F, H and IR\,6}
Both HH\,119\,E and F are previously undetected Herbig-Haro objects. As can be seen in the image, they both have separated nebulous features between
filters in the direction expected if they originate from the central outflow source. HH\,119\,F is seen as a H$\alpha$ bow shock, followed by a
similar S[II] feature. It resembles a smaller version of HH\,119\,B moving approximately ESE, as suggested by its structure and supported by its
proper motion.
HH\,119\,E is probably moving almost exactly to the east, as it is perfectly lined up with HH\,119\,A--D and the outflow source. As seen in the image for object E, and suggested by the optical spectra of other HH objects in this flow (B and C), object E must have very
strong [OI]~$\lambda\lambda$~6300/63 emission (as the leading shock is strongest in the $R$ band) followed by a H$\alpha$ shock.

There is another, although very faint, H$\alpha$ feature to the east of these objects in the HH flow but mid-way between them in declination.
This object, HH\,119\,H, is however clearly seen at 4.5\,$\mu$m using Spitzer (IR\,6 in Fig.\ref{B335_IRAC}). Details about this object are given
in Table\,\ref{shocksum}.

\subsubsection{HH119 G, IR\,2, 3 and 4}
In the blue-shifted lobe we find IR objects 2--4 moving as a group roughly to the ESE. Except for the brightest S(1) peak, IR\,4a (G), these objects
are only seen in the near-IR deep field. They all have slower projected velocities than the A--D optical bipolar flow, and especially features
3 and 4 have much slower space velocities. 

As suggested by the Spitzer image (Fig.\,\ref{B335_IRAC}) these shocks could be the result of turbulent interaction of molecular material from
the surrounding cloud and the outflow, near the cavity walls. This interaction could be what is slowing down the outflow and creating these shocks.
The reason why they are only seen in the near-IR could be a combination of extinction and the amount of H$_2$ relative
to atoms and ions (such as H, S+ and O). In a slow magnetically supported (C-type) shock, molecules are not as dissociated as in fast
(J-type) shocks. IR\,2 is moving much faster than  objects 3 and 4, in a direction more to the east, within the cone that has already been cleared
out by the outflow.

\subsection{HH119 I and the IR\,8 system}
In our 2007 images of western B335 we see a very likely counterflow (Fig.~\ref{New_S1_flow}) to the ESE flow, consisting of a system of at
least 15 H$_2$ knots. Careful image re-reduction of our optical deep field close to the mosaic edge, where the signal is lower due to few
overlaps and the flatielding is very sensitive to vignetting, revealed an optical (H$\alpha$ and [SII]) counterpart, HH\,119\,I (Fig.~\ref{New_opt_flow}), to one of these H$_2$ knots. This WNW counterflow (although no proper motions are known yet) is, similar to the ESE flow, much brighter in S(1) than
at H$\alpha$ or S[II] when compared to HH\,119\,A--F in the E--W flow. The detection of this system and the similar properties to the ESE flow opens
up the possibility that the outflow source may in fact be a binary source, with two slightly different oriented outflow axes giving rise to these
two flows.

\section{Summary and conclusions}

In an effort to investigate the Herbig-Haro flows in the nearby dark globule B335, a well-studied protostellar collapse candidate with three
previously known HH objects, we have taken optical spectra and observed an optical tri-colour deep
field (H$\alpha$ 6.7\,h, [SII] 6.2\,h and $R$ 39\,min) and a near-IR bi-colour deep field (2.12\,$\mu$m H$_2$ S(1) 6.5\,h and $K_S$ 48\,min), as
well as an additional near-IR field of western B335, using the 2.56\,m Nordic Optical Telescope and a
near-UV deep field ($U$ band 7.4\,h) with the 3.58\,m New Technology Telescope. In addition we present new
SPITZER\,/\,IRAC (3.5, 4.5 and 8.0\,$\mu$m) and MIPS (24\,$\mu$m) observations. The following results were obtained:

\vspace{3mm}
1. We discover five new HH objects (HH\,119\,D--H) in the blue-shifted (eastern) lobe and one new HH object (HH\,119\,I) in the red-shifted (western) lobe
of the outflow. Two of these HH objects (D and G) have previously observed near-IR counterparts (Hodapp~\cite{hodapp}). In our S(1) mosaic we detect
these as large extended shocks (object D being a large bow shock). In western B335 we also detect a system of at least 15 previously unknown H$_2$ knots.

2. By combining our observations with earlier epochs in the optical and near-IR we calculate proper motions of the previously known
HH objects (and the new HH\,119\,F) and 2.12\,$\mu$m H$_2$ S(1) knots using a time span of 15 and 9\,years, respectively.

3. The combined optical and near-IR proper motion maps, together with the deep images, show that the opening angle of the bipolar flow is already
very wide and reveals two sub-groups within the flow, a high velocity bipolar flow (space velocities 200--280\,km\,s$^{-1}$) in the E--W
direction (HH\,119\,A--F) that is bright in optical shock lines, and a much slower flow (15--75\,km\,s$^{-1}$) in the ESE direction within the blue
lobe (HH\,119\,G and IR\,2--4) that is bright in the S(1) line. We also propose a counterflow in a roughly WNW direction (red lobe), also
bright in S(1) and faint in H$\alpha$ and [SII].

4. All proper motion vectors originate from central B335. We suggest two possibilities, either the E--W and ESE--WNW shock groups belong to the same
outflow but are flowing in different parts of a cone-shaped cavity cleared out by the outflow, or the outflow source is binary in nature resulting
in two outflow axes with slightly different orientations.

5. Two of the new HH objects, HH\,119\,E/F, are equidistant to HH\,119\,A from the outflow source, while HH\,119\,C/D are roughly
equidistant to HH\,119\,B. The bipolar E--W flow is thus roughly symmetrical, with two outburst episodes suggested, 255 and 645 years ago, respectively, 
from proper motions. The conclusion that the central source has undergone several eruptive events has previously been suggested by
Reipurth et al. (\cite{reipurth}) who also used proper motions to estimate these two outburst ages (they estimated 350 and 850 years).

6. For HH\,119\,A, the proper motions of three H$\alpha$ condensations show that the shock is rapidly expanding, at a much faster rate than
would be expected from a linear expansion given its distance from the outflow source.

7. The [SII]\,/\,H$\alpha$ structure of the shocks indicate that the ESE group rams into a denser ambient medium, with the situation being
the opposite for the E-W flow.

8. Our optical spectra show HH\,119\,A to be of higher excitation than B and C (which are both low excitation objects).
We detect H$\beta$ for the first time in both A and B, making it possible to estimate the extinction towards these HH objects.
In stark contrast to a previous spectrum of HH\,119\,B, we detect strong [OI]~$\lambda\lambda$~6300/63 emission lines.
It is clear that object B must have changed a lot during the last decade or so.

9. We calculate the optical extinction towards HH\,119\,A and B to be A$_V$\,$\approx$\,1.4 and 4.4, respectively.

10. Using planar shock models we have estimated the shock properties of HH\,119\,A--C.
The resulting shock velocities are much lower (a factor 5--10) than the space velocities found through proper motions. This is however very
common for Herbig-Haro objects, and expected when gas with high space velocity is caught up by slightly faster gas, making the relative (shock) velocity
lower than the space velocities. This is also the case in bow shocks, where the shock velocity is lower in the wings than at the apex of the bow since
the projected velocity perpendicular to the shock is highest at the apex and lowest far out in the wings.
We find shock velocities of $\sim$60\,km\,s$^{-1}$ (A) and $\sim$35\,km\,s$^{-1}$ (B and C). This agrees well with object A being of higher
excitation, and B following in the wake of this object. The [SII]-weighted ionization fraction is also found to be much higher in object A. 

11. We detect HH\,119\,A, B and E in the $U$ band after background subtraction of scattered starlight throughout the globule. This emission is
proposed to be a combination of two spectral features, the [OII]\,$\lambda$3728 line and the blue continuum.

12. Most of the HH objects and S(1) knots are detected in the SPITZER\,/\,IRAC observations (channel 2, 4.5\,$\mu$m). An hour-glass shaped feature
is seen in the IRAC images, centred on the VLA outflow source, with a morphology that agrees with a cone-shaped E--W flow.
Even at 24\,$\mu$m it is unclear whether most of the light is direct or reflected, a single peak can be seen but it is offset by about
2\farcs8 to the SE from the VLA position.

\begin{acknowledgements}
	The Swedish participation in this research is funded by the Swedish National Space Board.
	This publication made use of the NASA/IPAC Infrared Science Archive, which is operated by the Jet Propulsion
	Laboratory, California Institute of Technology, under contract with the National Aeronautics and Space
	Administration, and data products from the Two Micron All Sky Survey, which is a joint project of the University
	of Massachusetts and the Infrared Processing and Analysis Center/California Institute of Technology, funded
	by the National Aeronautics and Space Administration and the National Science Foundation.
	We would like to thank Klaus-Werner Hodapp, Institute for Astronomy, University of Hawaii for providing the
	2.12\,$\mu$m H$_2$ mosaic we used as the first epoch in our proper motion calculations. We would also like to
	thank Bo Reipurth, Institute for Astronomy, University of Hawaii for providing us with the first epoch H$\alpha$ images we
	used in our optical proper motion calculations. Also, we want to thank Sven Olofsson, Stockholm Observatory, for providing
	us with the $U$ band image.
\end{acknowledgements}


\begin{thebibliography}{}

\bibitem[1993]{andre} Andr\'{e} P., Ward-Thompson D., Barsony M., 1993, ApJ 406, 122
\bibitem[1992]{anglada} Anglada G., Rodr\'{\i}guez L.F., Cant\'{\o} J., et al., 1992, ApJ 395, 494
\bibitem[2001]{avila} Avila R., Rodr\'{\i}guez L.F., Curiel S., 2001, Rev.\,Mexicana, Astron. Astrofis. 37, 201
\bibitem[1989]{cardelli} Cardelli J.A., Clayton G.C., Mathis J.S., 1989, ApJ 345, 245
\bibitem[1990]{chandler} Chandler C.J., Gear W.K., Sandell G., et al., 1990, MNRAS 243, 330
\bibitem[1982]{frerking} Frerking M.A., Langer W.D., 1982, ApJ 256, 523
\bibitem[1987]{frerking87} Frerking M.A., Langer W.D, Wilson R.W, 1987, ApJ 313, 320
\bibitem[2005]{notcamdist} G\aa lfalk M., 2005, NOT Annual report 2004, p18-19
\bibitem[2007]{galfalk06} G\aa lfalk M., Olofsson G., 2007, A\&A 466, 579
\bibitem[1994]{hartigan94} Hartigan P., Morse J.A., Raymond J., 1994, ApJ 436, 125
\bibitem[1999]{hartigan99} Hartigan P., Morse J.A., Tumlinson J., et al., 1999, ApJ 512, 901
\bibitem[2006]{harvey06} Harvey P.M., Chapman N., Lai S-P., et al., 2006, ApJ 644, 307
\bibitem[2001]{harvey01} Harvey D.W.A., Wilner D.J., Lada C.J, et al., 2001, ApJ 563, 903
\bibitem[2003]{harvey03} Harvey D.W.A., Wilner D.J., Myers P.C., 2003, ApJ 583, 809
\bibitem[1998]{hodapp} Hodapp K.W., 1998, ApJ 500, L183
\bibitem[1996]{kaufman} Kaufman M.J., Neufeld D.A., 1996, ApJ 456, 611
\bibitem[1993]{keenan} Keenan F.P., Hibbert A., Ojha P.C., et al., Phys. Scripta 48, 129
\bibitem[1983]{keene} Keene J., Davidson J.A., Harper D.A., et al., 1983, ApJ 274, L43
\bibitem[1994]{morse94} Morse J.A., Hartigan P., Heathcote S., et al., 1994, ApJ 425, 738
\bibitem[1993]{morse93} Morse J.A., Heathcote S., Cecil G., et al., 1993, ApJ 410, 764
\bibitem[2004]{noriega04} Noriega-Crespo A., Morris P., Marleau F.R., et al., 2004, ApJS 154, 352
\bibitem[1996]{ramsbottom} Ramsbottom C.A., Bell K.L., Stafford R.P., 1996, Atomic Data and Nuclear Data Tables 63, 57
\bibitem[1983]{reipurth83} Reipurth B., 1983, A\&A 117, 183
\bibitem[2002]{reipurth02b} Reipurth B., Heathcote S., Morse J., et al., ,2002, AJ 123, 362
\bibitem[1992]{reipurth} Reipurth B., Heathcote S., Vrba F., 1992, A\&A 256, 225
\bibitem[2002]{reipurth02} Reipurth B., Rodr\'{\i}guez L.F., Anglada G., Bally J., 2002, AJ 124, 1045
\bibitem[2005]{smith} Smith M.D., Rosen A., 2005, MNRAS 357, 1370
\bibitem[1979]{tomita} Tomita Y., Saito T., Ohtani H., 1979, PASJ 31, 407
\bibitem[1986]{vrba} Vrba F.J., Luginbuhl C.B., Strom S.E., Strom K.M., et al., 1986, AJ 92, 633

\end{thebibliography}
\end{document}